\documentclass[]{aastex701}

\newcommand{\axaf}{\mbox{\em Chandra\/}}

\accepted{\today}

\shorttitle{\axaf\ observation of the PKS 0637$-$752 jet}
\shortauthors{Maithil et al.}

\begin{document}

\title{Chandra X-ray imaging and IC/CMB model for the inner jet of PKS 0637$-$752}

\author[0000-0002-4423-4584]{Jaya Maithil}
\affiliation{Center for Astrophysics $\vert$ Harvard \& Smithsonian, 60 Garden St, Cambridge, MA 02138, USA}
\email[show]{jaya.maithil@cfa.harvard.edu}

\author[0000-0001-8252-4753]{Daniel A. Schwartz}
\affiliation{Center for Astrophysics $\vert$ Harvard \& Smithsonian, 60 Garden St, Cambridge, MA 02138, USA}
\email{dschwartz@cfa.harvard.edu}

\author[0000-0002-0905-7375]{Aneta Siemiginowska} 
\affiliation{Center for Astrophysics $\vert$ Harvard \& Smithsonian, 60 Garden St, Cambridge, MA 02138, USA} 
\email{asiemiginowska@cfa.harvard.edu}

\author[0000-0002-1516-0336]{Diana M. Worrall} 
\affiliation{H.H. Wills Physics Laboratory, University of Bristol, Tyndall Ave., Bristol BS8 1TL, UK} 
\email{d.worrall@bristol.ac.uk}

\author[0000-0003-3203-1613]{Preeti Kharb}
\affiliation{National Centre for Radio Astrophysics, Pune, Maharashtra 411007, India}
\email{preeti.kharb@gmail.com }



\begin{abstract}
We present the X-ray surface brightness maps of the jet from the quasar PKS 0637$-$752, using two epochs of archival Chandra observations (1999 and 2017). We present, for the first time, a model of the faint inner jet extending from 3.4 to 7 arcsec from the core, interpreting its X-ray emission as inverse Compton scattering of cosmic microwave background photons by the synchrotron-emitting relativistic electrons. The inner jet contrasts with the bright outer part of the jet, dominated by several discrete knots for which upper limits to the Fermi gamma-ray flux rule out the inverse Compton mechanism. It is the first detailed inverse Compton model published for this inner jet. We examine three physically motivated scenarios, distinguished by the high-energy cutoff of the electron distribution, that can account for the observed radio, millimeter/submillimeter, and X-ray emission while remaining consistent with optical and gamma-ray upper limits. We also report for the first time X-rays from the radio lobe at the end of the receding jet.
\end{abstract}

\keywords{Relativistic jets --- Radio loud quasars --- X-ray quasars}


\section{Introduction} \label{sec:intro}

The exploration of astrophysical phenomena at X-ray wavelengths has been fundamentally transformed by the \axaf\ X-ray Observatory \citep{Weisskopf2000}, enabling us to image celestial objects with unprecedented spatial resolution \cite[e.g., see science reviews by][]{Wilkes2019}. The very first target to which \axaf\ was pointed was the quasar PKS 0637$-$752 ($z=0.651$, \citealt{Savage1976}). It was known to be an X-ray point source \citep{Elvis1984}  and thus planned for the initial operations to find the best focus position for the X-ray cameras. Surprisingly, these observations revealed a bright X-ray jet \citep{Schwartz2000,Chartas2000}, well resolved and clearly associated with the kilo-parsec-scale radio jet known for this object \citep{Tingay1998}. Due to the exquisite angular resolution, this extended jet did not compromise the use of PKS 0637$-$752 to successfully focus the instruments. 

Until \axaf, the study of jets had been primarily within the domain of radio observations. Anticipated on theoretical grounds even prior to their empirical detection~\citep{Rees1971}, jets served as a conceptual framework to elucidate how substantial energy is ferried to the distant lobes of radio galaxies. They have a profound imprint on the energy budget of Active Galactic Nuclei (AGNs), with the central black hole serving the role as the ultimate engine~\citep{Meier2003}. \axaf\ showed that the enthalpy flux carried by jets solves the problem of how the collapse of clusters of galaxies is prevented~\citep{Fabian2000}.  It is evident that jets are part of the feedback process that correlate the mass of the central supermassive black hole with the bulge mass of the galaxy~\citep{Fabian2012,Mukherjee2025}. 
Observations of jets are important for learning about the processes of electron acceleration and the physics of how the jet is accelerated and collimated. The spectral energy distribution (SED) across a broad bandwidth reveals the distribution of relativistic electrons in different energy ranges \citep[see e.g.,][]{Harris2006,Blandford2019}. 

The X-ray discovery papers could only make models of the jet X-ray emission by invoking  complex multi-component emission regions, ad hoc magnetic field variations or coincidental bending of the jet about an axis in the plane of the sky. The seminal works of \citet{Tavecchio2000} and \citet{Celotti2001} solved this dilemma by invoking bulk relativistic motion of the jet. In such a case a single population of relativistic electrons could produce synchrotron emission at GHz frequencies, while the extension of that population to lower energies could produce inverse Compton X-rays via scattering of the cosmic microwave background\footnote{We refer to this as the IC/CMB mechanism, for short.}. This mechanism was subsequently widely invoked to explain X-ray jets in powerful quasars and FR II radio sources \citep[e.g.][]{Siemiginowska2002,Siemiginowska2003,Marshall2005,Cheung2006, Simionescu2016,Harris2017,Worrall2020,Migliori2022, Maithil2025}. Interestingly, no detailed model has ever been made for the entire PKS 0637-752 jet, although parameters for the bright knots at distances 7 to 11 arcsec from the quasar were presented by \citet{Schwartz2005}\footnote{Available at www.slac.stanford.edu\/econf\/C041213\/proc\/sess\_5.htm, see also \citet{Schwartz2007}} and for the knot at 7\farcs8 by \cite{Mehta2009}.

Despite the widespread application of the IC/CMB model, it clearly is not a universal explanation for even the quasar X-ray jets \citep[e.g.,][]{Hardcastle2006,Jester2006,Siemiginowska2007}.  Specifically, the expectation that the X-ray to radio flux ratio should increase as (1+z)$^4$ \citep{Schwartz2002}, was discrepant with observations by \citet{Marshall2011}. Furthermore, the GHz-emitting electrons should primarily lose their energy via IC/CMB and radiate in the 100 GeV to TeV range.
Fermi $\gamma$-ray upper limits effectively ruled out this possibility for the bright knots of PKS 0637$-$752 \citep{Meyer2015,Meyer2017}. Further investigations have found similar results for other quasar jets \citep{Breiding2023}, and an indication of general X-ray time variability of jet knots \citep{Hardcastle2016,Meyer2023} shows they must be emitting synchrotron radiation from short-lived TeV range electrons. 

The TeV electrons responsible for emitting synchrotron X-rays have lifetimes of only a few hundred years \citep{Harris2002}. Such electrons would have to be accelerated at hundreds of independent locations spaced on the order of every 100 pc to produce the apparently continuous X-ray jets. To avoid IC/CMB, the jets must be at most mildly relativistic. Thus the radiation we measure is not highly boosted, and it is not clear that such jets can efficiently transport enthalpy flux to do work on intracluster gas or to power radio lobes.

Our work aims to reconcile the synchrotron vs. IC/CMB dichotomy for the case of an X-ray jet associated with a flat spectrum quasar PKS 0637-752 \citep{Healey2007}. We present a semi-analytical model using synchrotron and IC/CMB mechanisms to explain the multi-wavelength emission from the inner 3.4–7 arcsec region of the jet. Following \citet{Murphy1988} as implemented in \citet{Maithil2025}, we adopt a prior to break the degeneracy between the bulk Lorentz factor and Doppler boosting factor. Section \ref{sec:multi} presents the two epochs of X-ray observations and accompanying multi-wavelength data. The model is briefly described in Section \ref{sec:model}, with full details in the Appendix \ref{app:iccmb}. Section \ref{sec:results} presents results for three possible scenarios that differ primarily in their electron energy cutoffs. Section \ref{sec:discussion} provides the discussion and conclusions. We adopt a cosmology with ${\rm H}_{0}=67.8\rm\,km\,s^{-1}\,Mpc^{-1}$, $\Omega_{\rm
M}=0.308$ and $\Omega_{\rm \Lambda}=0.692$, \citep{Planck2016A&A...594A..13P}

\section{Multi-wavelength Dataset and Analysis}
\label{sec:multi}
We present analysis and modeling of two epochs of \axaf\, observations of PKS 0637$-$752: the first epoch of 6 sets from 1999 calibration observations in Sec.~\ref{sec:1999}, and the second one of 5 sets from 2017 pointed observations in Sec.~\ref{sec:followup}. Archival radio and millimeter/submillimeter data from the Australia Telescope Compact Array (ATCA) and Atacama Large Millimeter/submillimeter Array (ALMA) are described in Sec.~\ref{sec:atca} and Sec.~\ref{sec:alma} respectively. We introduce optical Hubble Space Telescope (HST) and $\gamma$-rays Fermi limits in Sec.~\ref{sec:fermi}. 

\subsection{X-ray calibration observations, 1999 }
\label{sec:1999}

During the August 14-20, 1999, initial \axaf\, focusing operations PKS 0637-752 was observed 25 times with different configurations of ACIS (Advanced CCD Imaging Spectrometer), and different focus and offset pointing parameters. \citet{Chartas2000} used 23 of those that totalled 108\,ks to study the flux and spectrum of the quasar and prominent western jet. \citet{Schwartz2000} used six observations (available at \dataset[doi.org/10.25574/cdc.277]{https://doi.org/10.25574/cdc.277}) totaling 34.1\,ks when ACIS-S3 (back side illuminated CCD) was within $\pm$0.25 mm of best focus. The ACIS focal plane temperature was at -100\,C, a relatively warm temperature that has 
not been used for observations during the science mission and that
never has been re-calibrated on-orbit. Nonetheless, these observations took place before significant molecular contamination built up on the optical blocking filter\footnote{see ACIS section in \url{https://cxc.cfa.harvard.edu/proposer/POG/}}, and also before most of the charge transfer inefficiency degradation due to soft proton exposure. They therefore offer a unique opportunity to observe the $< 1$ keV photons from this system \citep{Mueller2009}. 

The original publication of these 1999 data \citep{Schwartz2000} used H$_0=50\rm\,km\,s^{-1}\,Mpc^{-1}$. With the updated cosmology parameters used here, the original luminosity of the total jet would be reduced to 2.4$\times$ 10$^{44}$erg s$^{-1}$, under the assumption that it is isotropic.
\begin{figure*}[h!]
\centering
\includegraphics[width=0.65\columnwidth,trim=0.0cm 0cm 0.0cm 0.cm,clip]{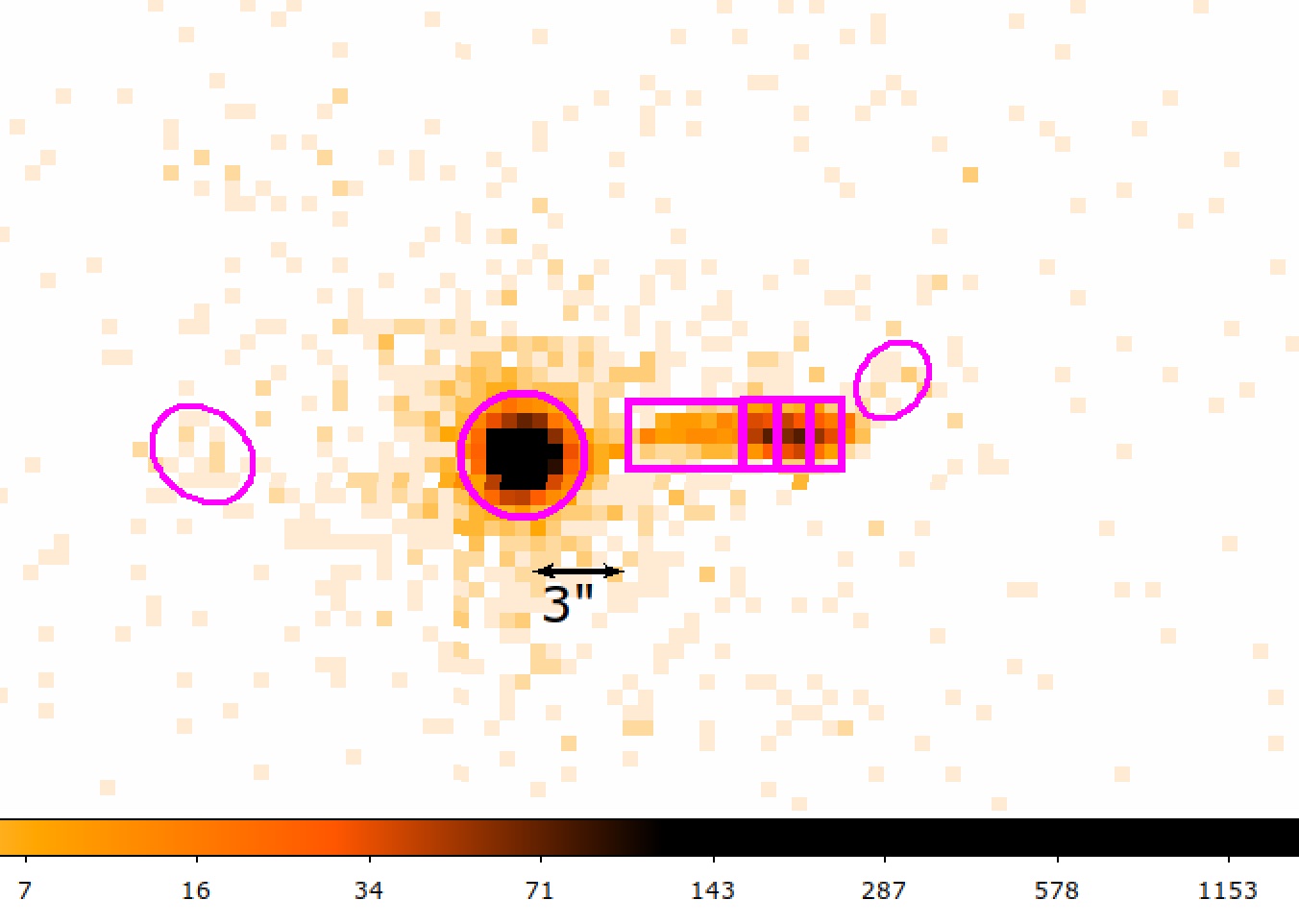}
\caption{\small The 34.1 ks image of PKS 0637$-$752 from the August 1999 calibration observations that were within 0.25 mm of the best focus. 
    \axaf\  ACIS-S3 data from 0.3--7 keV is binned in 0\farcs492 native ACIS pixels. The X-ray image is displayed in a logarithmic scale, and the color bar indicates counts per pixel. The faintest orange is one count, and white are pixels with zero count. Background is 0.051 counts per pixel. A 2\farcs5 radius circle is drawn around the quasar. The 3\farcs6 $\times$ 2\farcs2 rectangle shows our definition of the inner jet as used in our analysis. The three 1\arcsec $\times$ 2\farcs2 boxes include the bright knots in the outer jet. The two ellipses with major axis at angles 50\degr and 321\degr east of north, respectively, define the east and west X-ray ``lobes," respectively. Details of the regions are also listed in Table \ref{tab:regions}.}
\label{ChandraInnerJetCommissioning}
\end{figure*}

Fig.~\ref{ChandraInnerJetCommissioning} shows the near-focus data from the initial calibration observations in 1999. In this paper, we analyze and publish for the first time the X-ray data from the inner jet region, (the larger rectangle), and from the eastern radio lobe, (the larger ellipse to the east), and the western region where the X-ray jet bends, (the western ellipse). All regions are defined in  Table \ref{tab:regions}. In section \ref{sec:followup}
we compare these regions with the follow-up data from \axaf~Cycle 18 observations. 

Those three regions respectively contain 179, 17, and 17 X-ray photons in the 0.3 to 7 keV band. For spectral analysis of these data we use the ancillary response files (arf) and redistribution matrix functions (rmf) from \citet[][see their section 2]{Mueller2009}. These were made with the tools that weight the calibration products for extended sources. They used the -100\degr C `Fits Embedded Function' files developed prior to launch. The Chandra X-ray Center has not refined or maintained calibration products for this temperature, so there may be uncertainties that are not quantifiable. We fit an absorbed power-law model assuming the frozen Galactic absorption of 9.22 $\times$ 10$^{20}$ H-atom cm$^{-2}$. The inner jet photon index is 1.95 $\pm$ 0.13, with an 0.5-7 keV energy flux of (2.21 $\pm$ 0.18) $\times$ 10$^{-14}$ erg s$^{-1}$ cm$^{-2}$. With large uncertainties, the two elliptical regions shown in Fig.~\ref{ChandraInnerJetCommissioning} have similar photon indices of 2.05$\pm$0.54 and 1.89$\pm$0.46 and an energy flux (2.25$\pm$0.65)  $\times$ 10$^{-15}$ erg s$^{-1}$ cm$^{-2}$. 

\begin{deluxetable}{crcccc}[h!]
\tablecaption{ Observation log for the \axaf\  data. \label{tab:table2}}
\tablehead{
\colhead{Reference} & \colhead{ObsID} & \colhead{Date} & \colhead{Live time} \\\colhead{} & \colhead{} & \colhead{} & \colhead{(ks)}\\
\colhead{(1)} & \colhead{(2)} & \colhead{(3)} & \colhead{(4)}
}
\startdata
1999 calibration & 62550    & 1999-08-15  &  5.20  \\
    { }	           & 62553    & 1999-08-15  &  4.88  \\
    { }	           & 62554    & 1999-08-14  &  11.22  \\
    { }	           & 473      & 1999-08-20  &  3.48  \\
    { }	           & 474	  & 1999-08-20  &  4.64	  \\
    { }	           & 475	  & 1999-08-20  &  4.64  \\
\hline
Cycle 18 & 19692    & 2017-11-27  &  25.35  \\
{ }	           & 20864    & 2017-11-27  &  21.61  \\
{ }	           & 20865    & 2017-11-28  &  21.21  \\
{ }	           & 20866    & 2017-11-29  &  22.54  \\
{ }	           & 20867    & 2017-11-30  &  22.49  \\
\enddata
\tablecomments{(1) Reference for the observation. (2)Observation Id. (3) Date of observation. (4) Exposure time.}
\end{deluxetable}

\begin{figure*}[h!]
\centering
\includegraphics[width=0.65\columnwidth,trim=0.0cm 0cm 0.0cm 0.cm,clip]{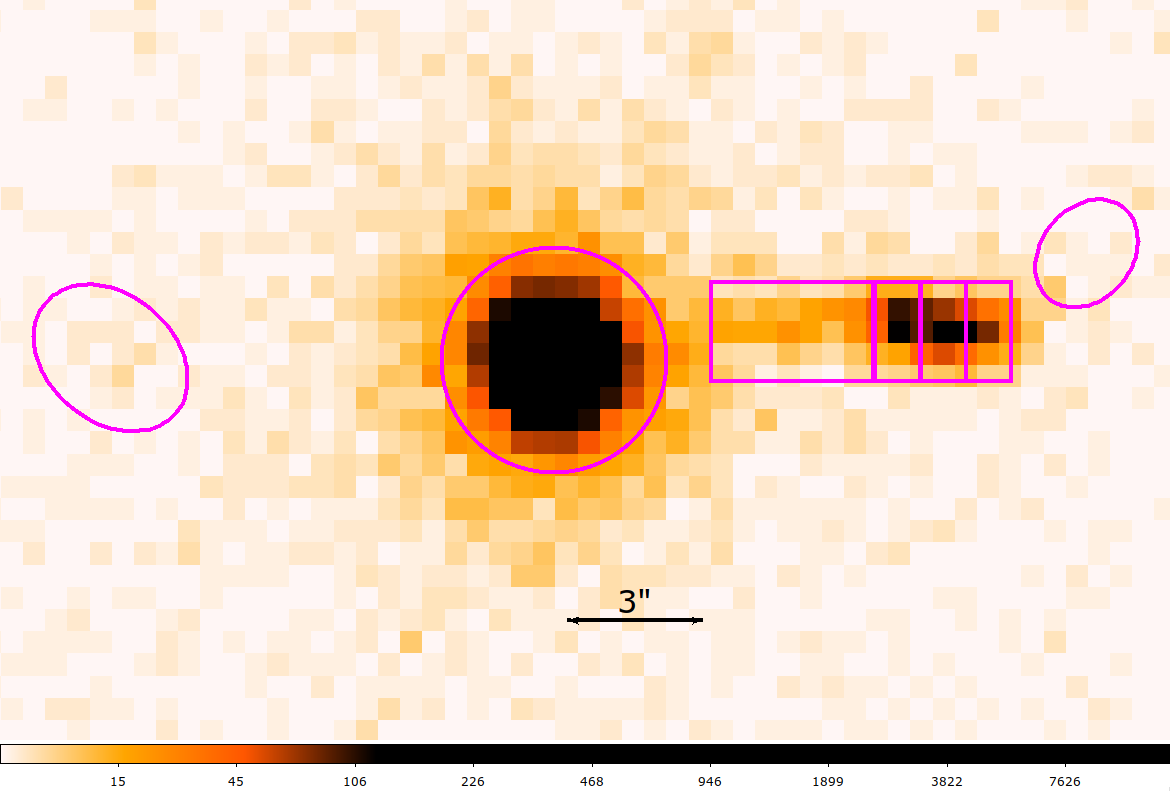}
\caption{\small The 113 ks merged \axaf\ ACIS-S image of PKS 0637$-$752 at 0.5-7 keV from 2017, binned in native ACIS pixels.  The X-ray image displayed in a logarithmic scale, and the color bar indicates counts per pixel. Regions defining the inner jet, the outer knots, and the two X-ray lobes are the same as in Fig.~\ref{ChandraInnerJetCommissioning} and are defined in Table \ref{tab:regions}}.
\label{image2017}
\end{figure*}

\subsection{\axaf~Cycle 18 observations \label{sec:followup}}

We present the five Cycle 18 ACIS-S observations of PKS 0637$-$752 (PI: Perlman; \dataset[doi.org/10.25574/cdc.411]{https://doi.org/10.25574/cdc.411}); a complete list is given in Table 1. The total exposure is 113.2 ks, 3.4 times deeper than calibration observations. We reprocessed each dataset with \texttt{chandra\_repro} (CALDB 4.10.2) to generate new level-2 event files. The quasar core centroid was measured in a $2 \farcs 5$ circular region using CIAO routine \texttt{dmstat}. The center of the X-ray source was then tied to the radio position (RA = 6:35:46.5240, DEC = -75:16:16.900) with \texttt{wcs\_update}, achieving sub-pixel alignment. Finally, the aligned event files were merged into a single dataset with CIAO routine \texttt{dmmerge}. The merged data is shown in Fig.~\ref{image2017}. 

\begin{figure}
    \centering
    \includegraphics[width=0.5\textwidth]{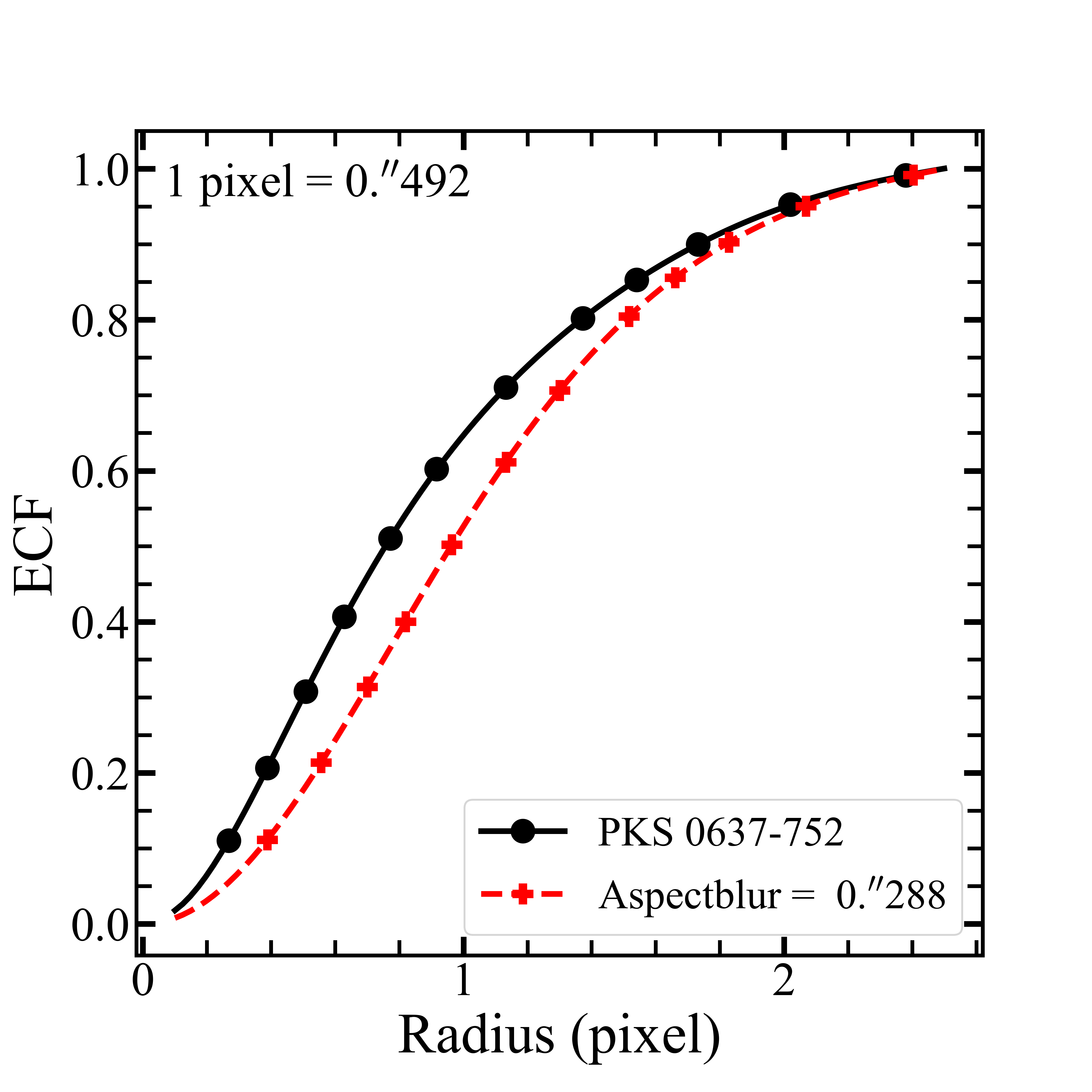}
    \caption{Enclosed count fraction as a function of the pixel radius. It compares the point spread function (PSF) of the observation (black) with the simulated PSF (red) using Aspectblur = $0 \farcs 288$.}
    \label{fig:ECF}
\end{figure}

Source and background spectra were extracted with the CIAO task \texttt{specextract}. The quasar core spectrum is extracted from a $0 \farcs 95$-radius circle centered on the radio position, enclosing 95\% of the counts. While the arcsecond–scale jet is  apparent in Fig.~\ref{image2017}, we quantified its statistical significance by comparing the data with detailed point-spread-function (PSF) simulations.  For each observation we generated 1,000 \texttt{SAOTrace} ray-tracing realizations based on the core spectrum and projected them onto the detector with \texttt{MARX}, employing the ACIS Energy-Dependent Subpixel Event Repositioning algorithm. Fig.~\ref{fig:ECF} shows that the simulated PSF using an \texttt{AspectBlur} value of~0.28 reproduces the observed PSF outside 1.5~pixels $\approx 0\farcs75$ while it diverges inside this radius. This level of agreement is sufficient to estimate how many quasar-core counts spill into the jet region and to correct the jet flux accordingly Obtaining the best imaging of the core is not our present goal.  Future studies that wish to isolate the X-ray emission of radio knots less than $2 \farcs 5$ from the nucleus could employ Bayesian image-analysis techniques such as \texttt{BAYMAX} \citep[e.g.,][]{Foord+2020}.

\begin{figure*}[h!]
\centerline{
\includegraphics[width=0.5\textwidth,trim=0 0 0 0]{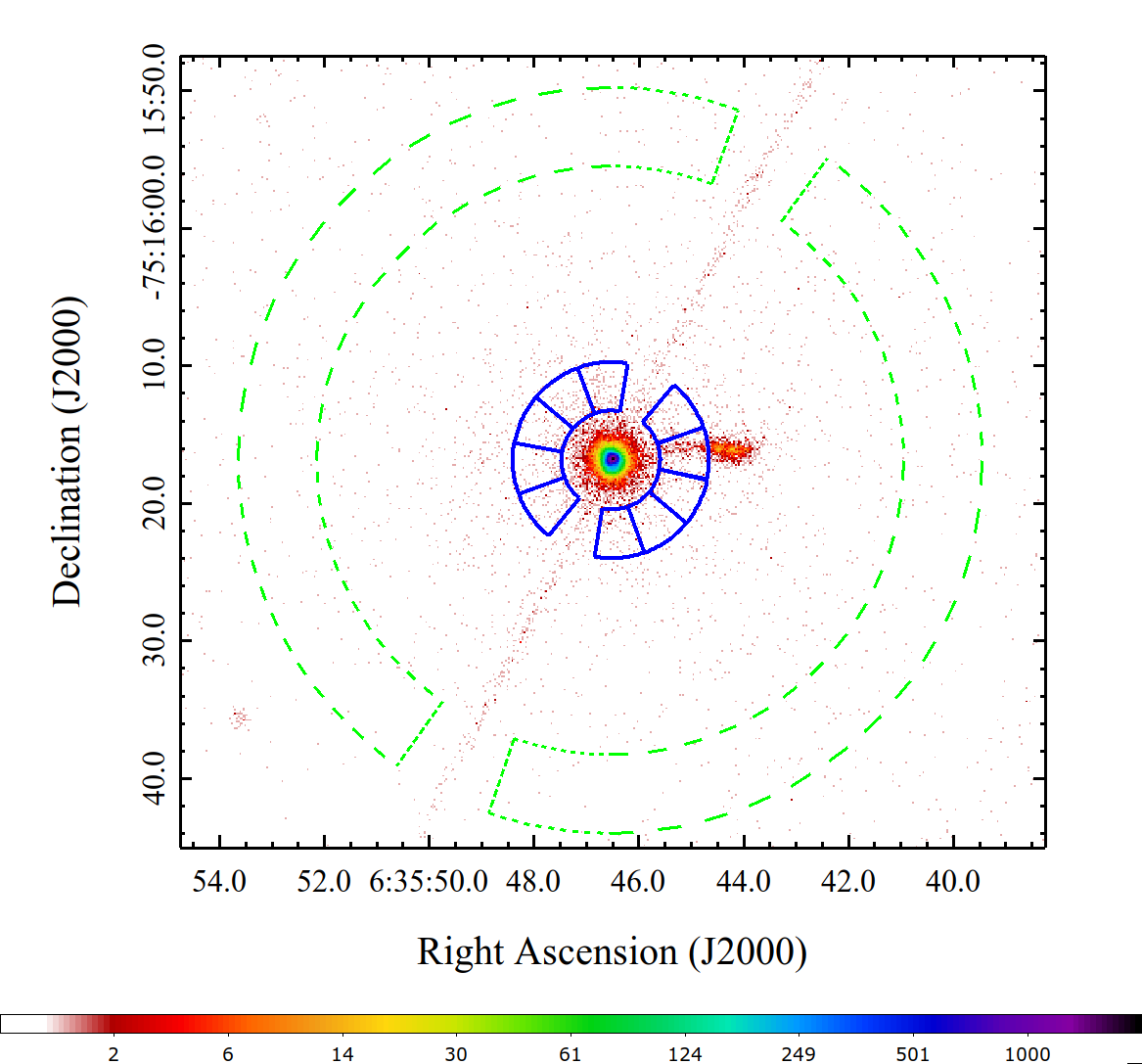}
\includegraphics[width=0.5\textwidth,trim=0 0 0 0]{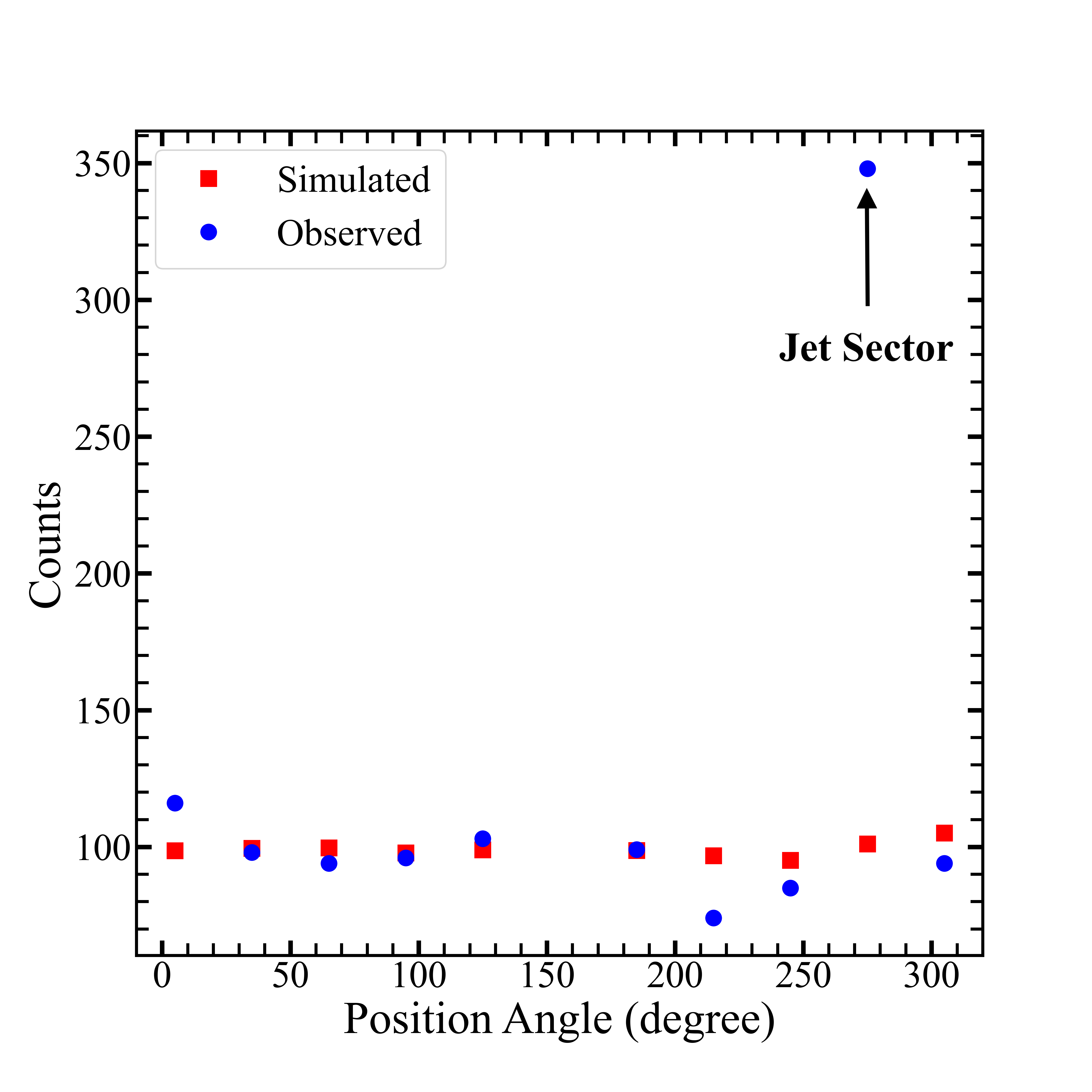}}
\caption{Left: ACIS-S image at 0.5-7 keV have been binned to 1/4 of a pixel (0\farcs  123) and overlayed with the region used to extract counts in the core and annular sectors at the extent of jet in blue. The sectors are at PA=(5, 35, 65, 95, 125, 185, 215, 245, 275, 305) and have inner and outer radius as $3 \farcs 6$ and $7 \farcs 1$. Background used is marked in green and contains 376 counts per pixel. Right: Observed net counts (blue) and simulated PSF counts (red) in the 0.5-7 keV energy band as a function of PA. Simulated counts in the sectors are normalized by the ratio of observed to simulated core counts. The jet sector marked with an arrow has 348 counts whereas the PSF simulation predicts 101 counts.}
\label{fig:CountsvsPA}
\end{figure*}

We tallied the X-ray counts within ten $30\degr$ annular sectors that span radii $3\farcs4$–$7\farcs0$ from the quasar, excluding the read-out streak (Fig.~\ref{fig:CountsvsPA}, left).  Treating the counts as Poisson variates, we derived the expected distribution due to scattering of the core X-rays for each sector from the PSF simulations, after normalizing the simulations to the observed core counts. Fig.~\ref{fig:CountsvsPA} (right) demonstrates that only the sector centered at PA~$= 275\degr$ exhibits a statistically significant excess, with $348$ detected counts versus $101$ predicted.  This comparison confirms that the visually obvious jet emission is indeed a robust, extended X-ray feature. The inner boundary of the jet box is set by the PSF wings of the bright quasar core. Between $2 \farcs 5$ and $3 \farcs 4$ the counts along the jet direction show no significant excess over any other position angles, so no jet emission is recoverable there.

\begin{figure*}[h!]
\centering
\includegraphics[width=0.9\columnwidth,trim=0.cm 0cm 0.0cm 0.cm,clip]{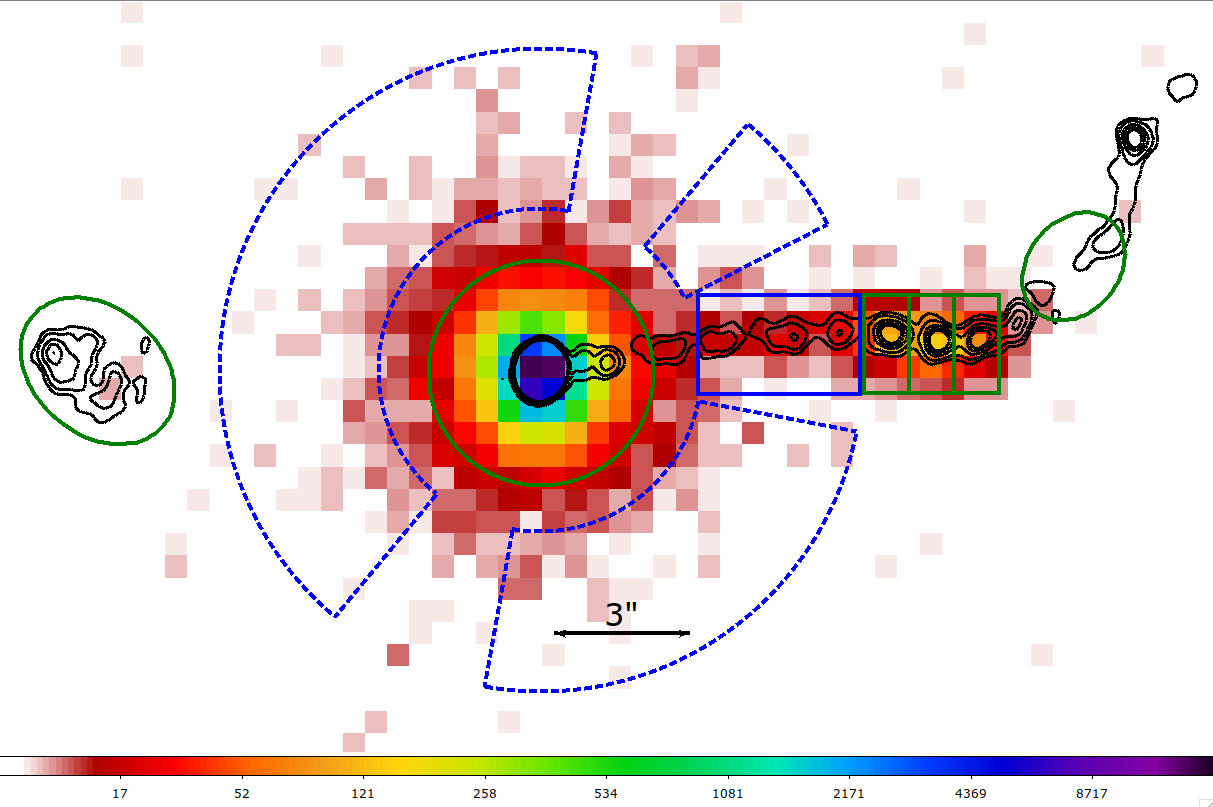}
\caption{\small \axaf ACIS-S image of PKS 0637$-$752 at 0.5-7 keV have been binned to 1 pixel ($0\farcs 492$). Black contours represent 20 GHz emission at ($-\sqrt{2}, \sqrt{2}, 2\sqrt{2}, 4\sqrt{2}, 6\sqrt{2}, 8\sqrt{2}) \times 0.6$ mJy beam$^{-1}$.
The jet region is marked by a $3 \farcs 6 \times 2 \farcs 2$ blue box, and its background by dotted pie sectors. The core, knots and lobes regions are also shown in green. All regions are defined in Table \ref{tab:regions}.}
\label{ChandraImage}
\end{figure*}

A $3 \farcs 6 \times 2 \farcs 2$ rectangle whose inner edge starts $3 \farcs4$ from the core center is used to extract the jet spectrum. This region defines the inner jet throughout the analysis. We use the annular sectors around the jet shown in Fig. \ref{ChandraImage} to extract the background spectrum. Three consecutive $1 \farcs 0 \times 2 \farcs 2$ boxes, the first beginning $7 \farcs 1$ from the core center, is used to extract the spectrum of the individual knots. These jet components show no significant emission beyond the 0\farcs9 radius enclosing 90\% of the PSF
counts, and we therefore treat them as unresolved in the direction perpendicular to the jet. The lobe and counter-lobe spectra are obtained from elliptical apertures: the western lobe uses semi-major and semi-minor axes of $1 \farcs 9$ and $1 \farcs 4$ at a position angle of 321\degr east of the north, and the eastern lobe uses axes of $1 \farcs 3$ and $1 \farcs 0$ at 50\degr. These regions are marked in Fig.~\ref{ChandraImage} and listed in Table \ref{tab:regions}. Background spectra for core, knots and lobes were extracted from two annular sectors centered on the core, with radii $21''–27''$ and position angles 63\degr–241\degr and 251\degr–414\degr, chosen to avoid the ACIS readout streak

\begin{table}
\centering
\caption{Regions used in the X-ray spectral fitting.}
\label{tab:regions}
\begin{tabular}{ll}
\hline\hline
Region & Definition \\
\hline
Core       &  circle(98.9438500, -75.2713611, 2\farcs5)\\
Inner jet          & box(98.9380656, -75.2711877, 3\farcs6, 2\farcs2, 360\degr) \\
Jet background     & pie(98.9438500, -75.2713611, 27, 50, 3\farcs4, 7\farcs0)\\
                   & +pie(98.9438500, -75.2713611, 80, 230, 3\farcs4, 7\farcs0)\\
                   & +pie(98.9438500, -75.2713611, 260, 349, 3\farcs4, 7\farcs0)\\
WK 7.8 & box(98.9354570, -75.2711873, 1\farcs0, 2\farcs2, 360\degr) \\
WK 8.9 &  box(98.9343708, -75.2711872, 1\farcs0, 2\farcs2, 360\degr)\\
WK 9.7 &  box(98.9332730, -75.2711880, 1\farcs0, 2\farcs2, 360\degr)\\
Lobe & ellipse(98.9309381, -75.2707063, 1\farcs0, 1\farcs3, 321)\\
Counter-lobe & ellipse(98.9545828, -75.2713497, 1\farcs4, 1\farcs9, 50) \\
\hline
\end{tabular}
\end{table}

Table 2 lists the net counts in the soft (0.5–2 keV), hard (2–7 keV), and broad (0.5–7 keV) bands and the resulting hardness ratio. Using Sherpa \citep{Freeman2001,Siemiginowska2024}, we simultaneously modeled the region and background spectra across 0.5–7 keV with a power law plus fixed Galactic absorption. Table 2 lists the resulting photon index, integrated 0.5–7 keV flux, and flux density at 1 keV for each region. In the 0.5–7 keV band the inner jet contains $282.7\pm16.9$ net counts, 1.57 times that of the 1999 calibration data, yet its spectral shape and brightness are statistically consistent, with a photon index of $\Gamma = 1.98\pm0.12$ and a flux of $(2.59\pm0.17)\times10^{-14}$ erg s$^{-1}$ cm$^{-2}$. The knot WK 7.8 has an X-ray spectrum similar to that of the inner jet, with a $\Gamma = 1.90\pm0.09$, whereas WK 8.9 and WK 9.7 are harder, with $\Gamma = 1.78\pm0.08$ and $\Gamma = 1.65\pm0.13$, respectively. The eastern and western lobes contain $29.3\pm5.7$ and $18.2\pm4.5$ net counts in the 0.5–7 keV band, respectively, and present similar photon indices of $\Gamma = 2.19^{+0.43}_{-0.40}$ and $\Gamma = 1.71^{+0.50}_{-0.47}$. Their corresponding energy fluxes (0.5–7 keV) are $(2.6\pm0.3)\times10^{-15}$ and $(1.7^{+0.6}_{-0.4})\times10^{-15}$ erg s$^{-1}$ cm$^{-2}$, entirely consistent with the 1999 measurements.

\begin{deluxetable*}{crrrccccc}[h!]
\tablecaption{ X-ray properties of the core, jet, knots, and lobes from Chandra Cycle 18 observations. \label{tab:table1}}
\tablehead{
\colhead{Component} & \multicolumn{3}{c}{Net Count} & \colhead{HR} & \colhead{$\Gamma_{\rm 0.5-7~  keV}$} & \colhead{$f_{\rm 0.5-7~  keV}$} & \colhead{$f_{\rm 1~  keV}$} \\
\cline{2-4}
\colhead{} & \colhead{{\footnotesize($\rm 0.5-7~  keV$)}} & \colhead{{\footnotesize($\rm 0.5-2~  keV$)}} & \colhead{{\footnotesize($\rm 2-7~  keV$)}} & \colhead{} & \colhead{} & \colhead{}\\
\colhead{(1)} & \colhead{(2)} & \colhead{(3)} & \colhead{(4)} & \colhead{(5)} & \colhead{(6)} & \colhead{(7)} & \colhead{(8)}
}
\startdata
Core    & $65462.6\pm 256.2$  & $35980.8\pm 190.0$ & $29770.2\pm172.8$ &  -0.09 &  1.50$\pm$0.01 & $663.88^{+3.12}_{-3.53}$ & $618.14^{+4.53}_{-4.35}$ \\
Jet box & 262.6$\pm$18.4 & 179.4$\pm$14.9 & 84.7$\pm$10.8 & -0.36 & 1.92$\pm$0.14 & 2.46$^{+0.17}_{-0.19}$ & 3.29$^{+0.34}_{-0.36}$ \\
WK 7.8   & $462.0\pm21.5$      & $287.6\pm17.0$     & $177.4\pm13.3$    &  -0.24 &  $1.90\pm0.09$ & $4.30^{+0.46}_{-0.41}$   & $5.70^{+0.42}_{-0.40}$\\
WK 8.9   & $517.0\pm22.8$      & $317.6\pm17.8$     & $201.4\pm14.2$    &  -0.22 &  $1.78\pm0.08$ & $4.95^{+0.49}_{-0.46}$   & $5.98^{+0.42}_{-0.38}$\\
WK 9.7   & $209.0\pm 14.5$     & $125.6\pm 11.2$    & $83.4\pm9.2$      &  -0.20 &  $1.65\pm0.13$ & $2.05^{+0.38}_{-0.30}$   & $2.20\pm0.24$  \\
All Knots   & $ 1230.0\pm35.1$    & $756.8\pm27.5$     & $478.1\pm21.9$    &  -0.23 &  $1.93\pm0.11$ & $9.91^{+0.59}_{-0.57}$   & $13.50^{+0.95}_{-0.92}$\\
Western Lobe    & $ 18.2\pm4.5$    & $12.3\pm3.6$     & $5.91\pm2.6$    &  -0.35 &  $1.71^{+0.50}_{-0.47}$ & $0.17^{+0.06}_{-0.04}$   & $0.20^{+0.08}_{-0.07}$\\
Eastern Lobe  & $29.3\pm5.7$    & $17.6\pm4.4$     & $11.7\pm3.7$    &  -0.20 &  $2.19^{+0.43}_{-0.40}$ & $0.26\pm0.03$   & $0.42^{+0.12}_{-0.11}$
\enddata
\tablecomments{(1) Component name. (2) Background-subtracted counts over the 0.5-7 keV energy band. (3) Background-subtracted counts over the 0.5-2 keV energy band. (4) Background-subtracted counts over the 2-7 keV energy band. (5) Hardness ratio, HR = (Hard-Soft)/(Hard+Soft). (6) Power-law photon index 0.5-7 keV energy band. (7) Flux in the 0.5-7 keV energy band in units of $\rm 10^{-14}~ erg~ s^{-1}~ cm^{-2}$. (8) Flux density at 1 keV in the units of $\rm 10^{-32}~ erg~ s^{-1}~ cm^{-2}~ Hz^{-1}$.}
\end{deluxetable*}

\newpage

\subsection{Archival ATCA imaging}\label{sec:atca}

PKS~0637$-$752 has been observed multiple times with ATCA.  For our analysis, we chose the highest resolution continuum images at 20.2, 17.7, 8.6 and 4.8 GHz frequencies. The 20.2 and 17.7 were observed in 2004 (Observer: L.\ E.\ H.\ Godfrey), and the 8.6 and 4.8 GHz in 1999 (Observer: R.\ Ojha). For each dataset, the raw visibilities were flagged and calibrated with MIRAD using standard procedures and imaged using with DIFMAP to produce the Strokes I maps used here. The elliptical restoring beams are $0\farcs42\times0\farcs35$ at 20.2 GHz, $0\farcs47\times0\farcs41$ at 17.7 GHz and $0\farcs8$ and $1\farcs6$ at 8.6 and 4.8 GHz. The corresponding theoretical RMS noise estimates are 0.125, 0.092, 0.096, and 0.102 mJy beam$^{-1}$ at 20.2, 17.7, 8.6, and 4.8 GHz, respectively.

\begin{figure*}[h!]
\centering
\includegraphics[width=0.45\columnwidth,trim=0.0cm 0cm 0.0cm 0.cm,clip]{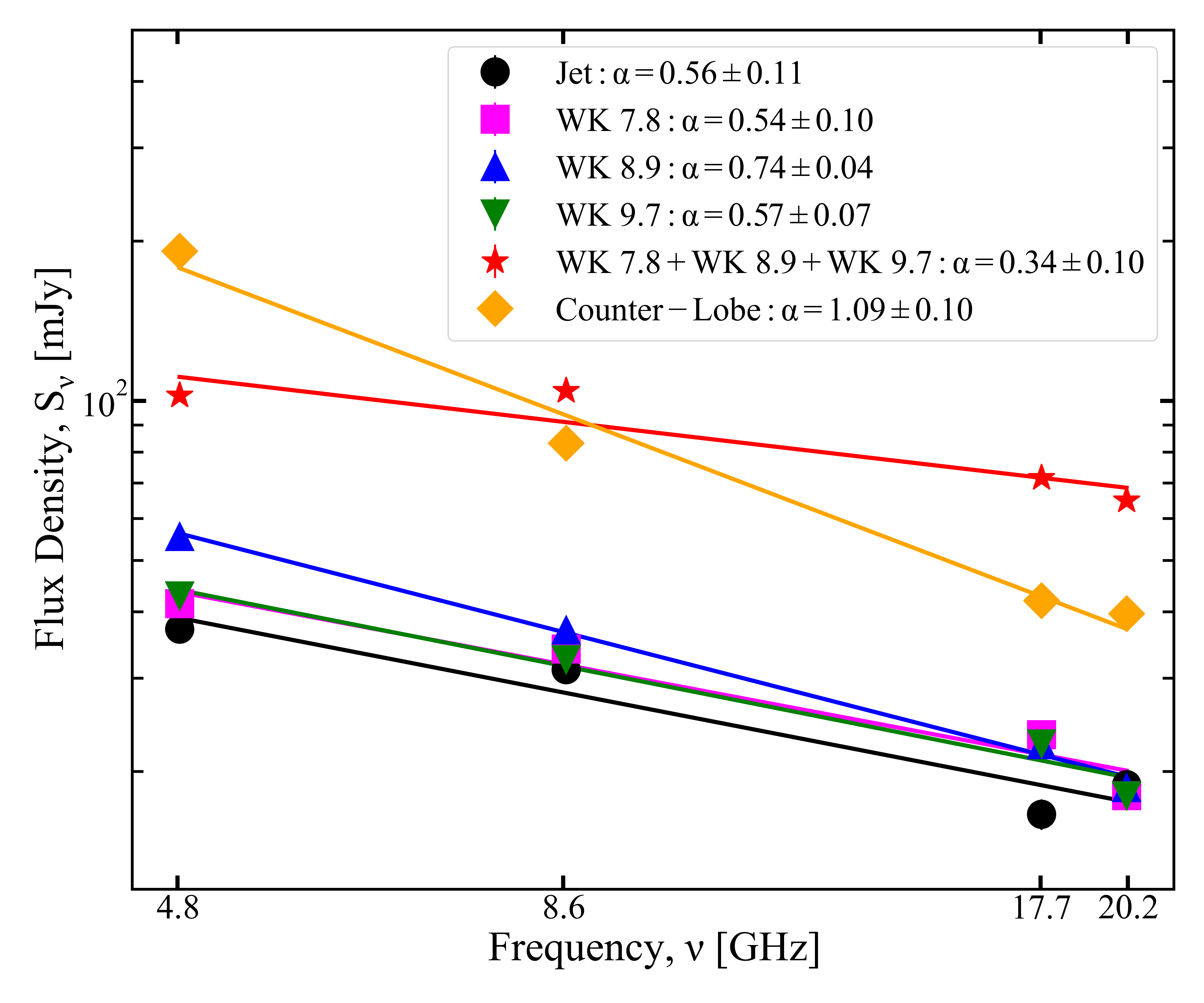}
\includegraphics[width=0.45\columnwidth,trim=0.0cm 0cm 0.0cm 0.cm,clip]{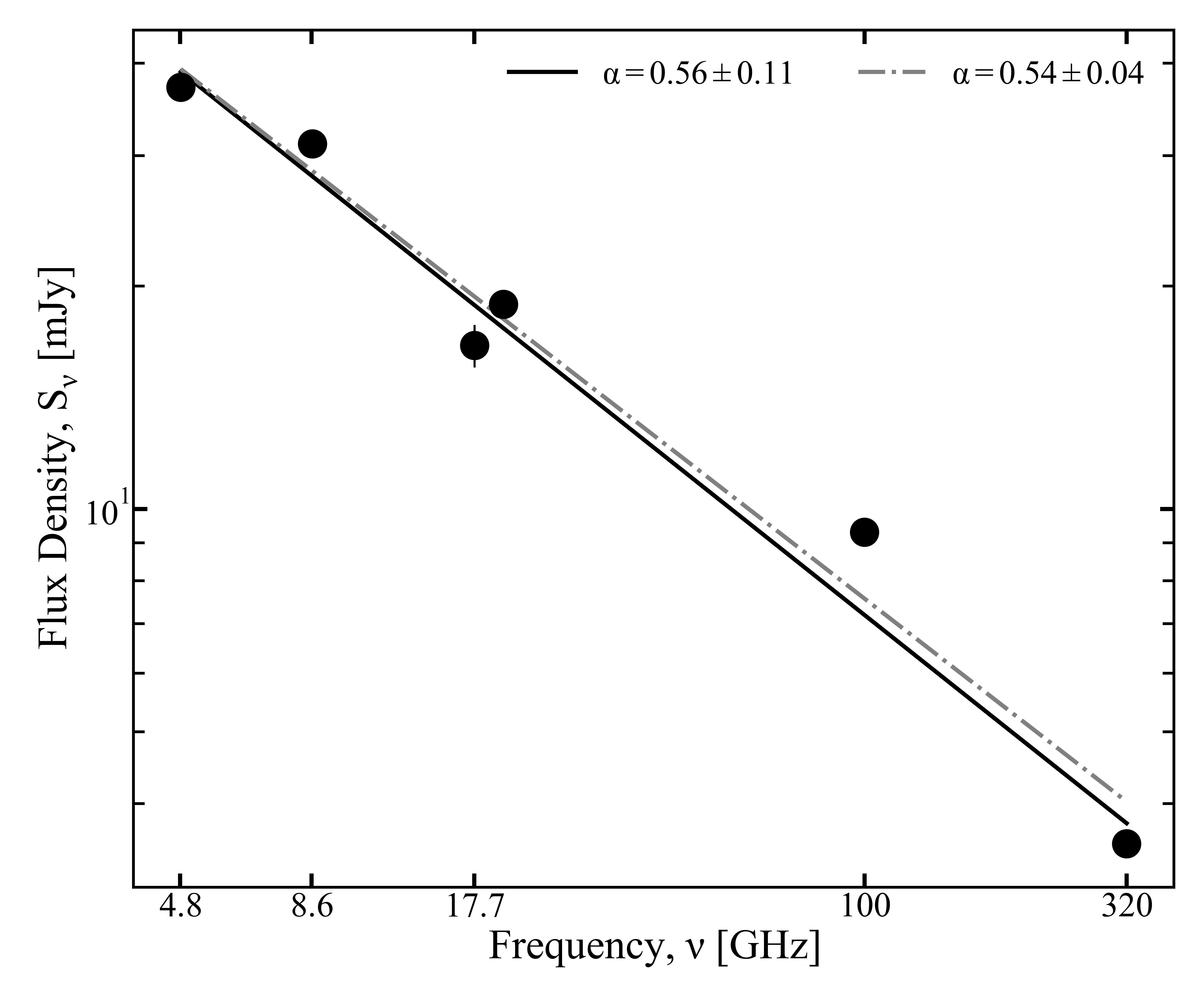}
\caption{Left: Radio spectral energy distribution of different components. Right: Spectrum of the inner jet including ALMA data.}
\label{fig:RadioSED}
\end{figure*}

Using \textsc{casa} task \texttt{imfit} we measured integrated flux densities for the unresolved inner jet segment, three brightness peaks along the jet (knots \textit{WK 7.8, WK 8.9, WK 9.7}), and the counter-lobe. We then fitted a simple power law, $S_{\nu}\propto\nu^{-\alpha}$, to the fluxes of each region, allowing a consistent determination of the radio spectral indices across nearly a factor of five in frequency. The resulting spectral indices are $\alpha_{\mathrm{jet}}=0.56\pm0.11$, $\alpha_{\mathrm{WK 7.8}}=0.54\pm0.10$, $\alpha_{\mathrm{WK 8.9}}=0.74\pm0.04$, $\alpha_{\mathrm{WK 9.7}}=0.57\pm0.07$, a combined‐knot slope of $\alpha_{\mathrm{WK 7.8 + WK 8.9 + WK 9.7}}=0.34\pm0.10$, and a markedly steeper counter-lobe value of $\alpha_{\mathrm{counter-lobe}}=1.09\pm0.10$. Fig.~\ref{fig:RadioSED} (left) illustrates these best–fitting power laws.

\begin{deluxetable*}{ccccccc}
\tabletypesize{\footnotesize}
\tablecaption{Radio Flux Densities by Component\label{tab:radio_flux}}
\tablehead{
\colhead{Frequency} & \colhead{Jet} & \colhead{WK 7.8} & \colhead{WK 8.9} & \colhead{WK 9.7} & \colhead{All Knot} & \colhead{Counter-Lobe} \\
\colhead{(GHz)} & \colhead{(mJy)} & \colhead{(mJy)} & \colhead{(mJy)} & \colhead{(mJy)} & \colhead{(mJy)} & \colhead{(mJy)}
}
\startdata
4.8  & 37.1$\pm$0.8 & 41.4$\pm$0.6 & 55.5$\pm$0.6 & 42.9$\pm$0.5 & 102.4$\pm$0.8 & 191.5$\pm$1.6 \\
8.6  & 31.1$\pm$1.2 & 34.1$\pm$0.7 & 37.0$\pm$0.7 & 32.5$\pm$0.6 & 104.5$\pm$1.1 & 83.2$\pm$1.2  \\
17.7 & 16.6$\pm$1.1 & 23.5$\pm$0.7 & 22.5$\pm$0.6 & 22.6$\pm$0.6 & 71.6$\pm$1.2  & 42.0$\pm$1.3  \\
20.2 & 18.9$\pm$0.8 & 18.0$\pm$0.5 & 18.7$\pm$0.5 & 18.0$\pm$0.5 & 64.9$\pm$1.0  & 39.7$\pm$1.2  \\
\enddata
\end{deluxetable*}

\subsection{ALMA millimeter and submillimeter imaging}\label{sec:alma}

We restrict our millimeter analysis to the inner jet which anchors the high‐frequency end of the synchrotron spectrum and provides critical constraints on the electron energy distribution in our model. PKS~0637$-$752 is a standard ALMA phase calibrator, and thus benefits from the exceptionally deep, high‐quality data collected by the \textsc{ALMACAL} survey \citep{Oteo+2016, Oteo+2017}. We requested the same Band~3 ($\nu\!\approx\!320$\,GHz) and Band~7 ($\nu\!\approx\!100\,\mu$m) visibilities analyzed by \citet{Meyer2017}. The \textsc{ALMACAL} products we received had already passed through ALMA’s automated calibration pipeline, yielding fully calibrated visibilities for every scheduling block; a brief round of self-calibration then refined the phase solutions (see \citealt{Oteo+2016} for details). 

We then subtract the variable core point source from each data set and combine all available visibilities, spanning multiple array configurations, using natural weighting. Primary beam corrected images reach rms sensitivities of $67.8\ \mu\mathrm{Jy\,beam^{-1}}$ (Band~3; $1\farcs61\times1\farcs12$) and $45.0\ \mu\mathrm{Jy\,beam^{-1}}$ (Band~7; $0\farcs32\times0\farcs23$). From these maps we measure inner-jet flux densities of $S_{100\,\mathrm{GHz}} = 9.30\pm0.07$ mJy and $S_{320\,\mathrm{GHz}} = 3.53\pm0.05$ mJy. Fig. \ref{fig:RadioSED} (right) shows that  fitting ATCA and ALMA flux densities together gives $\alpha_{\rm jet} = 0.54\pm0.04$ which is consistent with the spectral slope of the radio data alone ($0.56\pm0.11$). We adopt the synchrotron spectral index $\alpha_{\rm jet} = 0.56\pm0.11$ (electron energy index $p=2\alpha+1 = 2.12$) in our IC/CMB model. The ALMA measurements are used to constrain the maximum energy of the synchrotron emitting electrons.

\subsection{Optical, Infrared, and Fermi upper limits}
\label{sec:fermi}

Contemporaneous HST observations were carried out in November 2017 as part of the program GO 14696 (13 orbits; \citealt{Perlman2016hst}). The objective of that program was to measure the optical polarization of the jet. The polarization results have been presented by \citet{Perlman2020}. They detect polarized optical emission from the outer knots, while the inner jet falls below their detection limit. No optical polarization measurement therefore exists for the region we model in this work.

\begin{figure}[h]
\centerline{
\includegraphics[height=2.in]{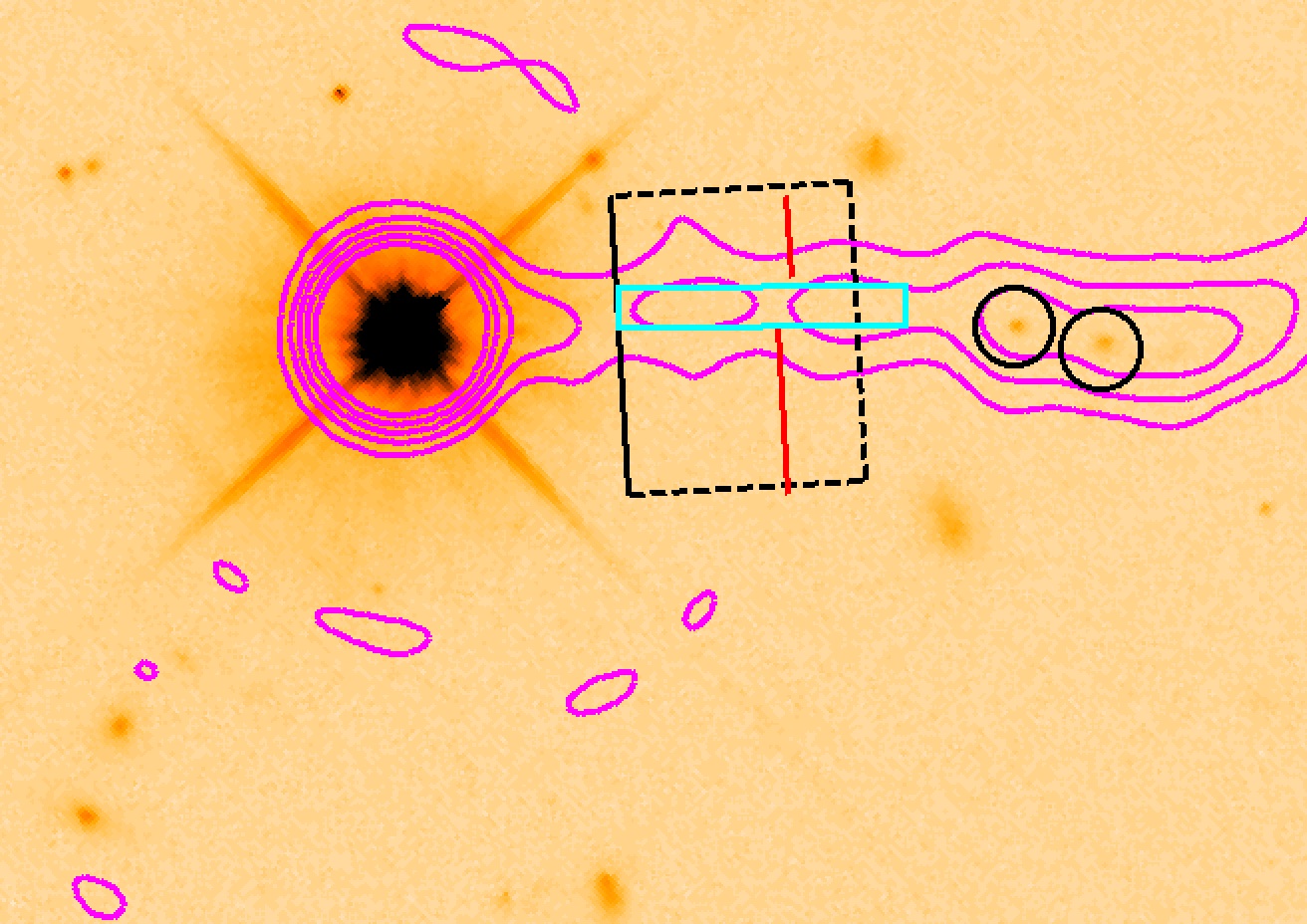}
\hspace{0.2in}
\includegraphics[height=2.in]{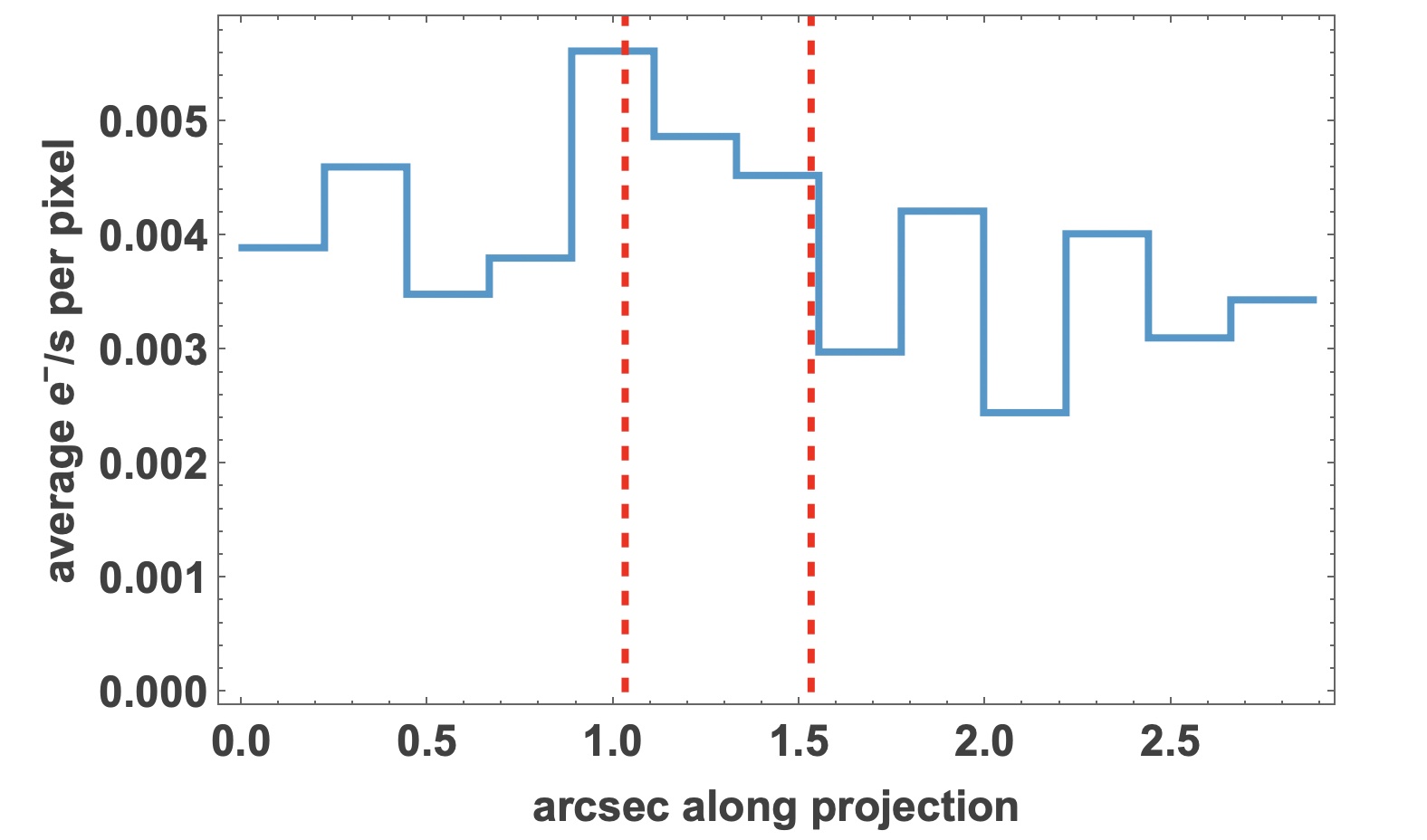}}
\caption{Left: HST ACS/WFC1 F606W image. Magenta contours show the 8.6 GHz contours of 1.7, 7.2, 19.1, 44.7, and 100 mJy $\rm beam^{-1}$. The cyan rectangle shows the inner jet region for which we obtain a limit to the flux at a frequency of $5.06 \times 10^{14}$ Hz. The black dashed rectangle shows the area used to sum all the signal in the jet and the background on either side, placed such that it avoids the bright background source. We project the signal within this area to the eastern edge marked by the solid black line. The solid red lines show the north to south direction of the projection, which corresponds to left to right in the right panel. Right: We divide the HST data inside the black dashed rectangle into 13 bins, each 0\farcs22 wide and parallel to the northern and southern sides of the rectangle. The blue trace shows the average count rate, in e$^-$$s^{-1}$ per 0\farcs05 ACS pixel, for each bin projected onto the solid black line. The dashed red lines at 1\farcs0 and 1\farcs5 delineate the boundaries of the putative optical emission from the inner jet shown as the cyan rectangle in the left panel.
\label{fig:hstimage}}

\end{figure}

Here we use just one of those archival observations, taken without a polarizing filter, to estimate a flux limit for the inner jet region. We use the data set JD7X04010, taken with the F606W filter on ACS/WFC1 (see \dataset[doi.org/10.17909/js73-am83]{https://doi.org/10.17909/js73-am83}). Fig.~\ref{fig:hstimage}, left, shows this HST image region around the quasar and inner jet, extending up to the bright jet knots. The optical image is aligned to the radio using the same procedure applied to the X-ray data. We tie the optical centroid of the core to its radio position. On the right panel, we show the projection of the HST data along lines parallel to the inner jet in 0\farcs22 $\times$ 3\farcs6 bins. The region of the inner jet is the 0\farcs5 $\times$ 3\farcs6 box 
(marked in cyan color) in the left panel aligned with the 8.6 GHz contours. The inner jet region is shown between the dashed red dashed lines in the right panel. We take the peak e$^-$s$^{-1}$ minus the mean of 0.0039, plus 3 times the standard deviation to give an upper limit optical flux = $1.6\times 10^{-16} \rm erg~ cm^{-2}~ s^{-1}$ in a 672\AA\ photometric bandwidth of the F606W filter around 5922\AA = $5.06 \times 10^{14}$ Hz. 

\citet{Mehta2009} report an HST/NICMOS detection of $2.81\times 10^{-16}~ \rm erg~ cm^{-2}~ s^{-1}$ at $1.87 \times 10^{14}$ Hz from knots (5.7\arcsec and 6.3\arcsec from the core) that fall in our inner jet region. We also adopted their Spitzer upper limits of $2.59 \times 10^{-15}~ \rm erg~ cm^{-2}~ s^{-1}$ and $1.67 \times 10^{-15}~ \rm erg~ cm^{-2}~ s^{-1}$ at $5.17 \times 10^{13}$ Hz and $8.33 \times 10^{13}$ Hz, respectively.

We adopted the Fermi upper limits from Table 1 of \citet{Meyer2017} and summarize their procedure here. They combine only the quiet $\gamma$-ray periods of the source, improving the sensitivity while avoiding contamination from the bright variable quasar core. From 2008 to 2016, weekly LAT bins were ranked based on test statistics and progressively combined. Flux measurements or 95\% upper limits were computed across five energy bands, assuming a photon index of 2.7. The deepest limits were achieved after combining 31 to 60 bins for each band, resulting in a minimum flux upper limit of $9.45 \times 10^{-14} \rm erg~ s^{-1} ~ cm^{-2}$ in the $1-3$ GeV range.

\section{Synchrotron and IC/CMB model for inner jet}
\label{sec:model}

We present the first detailed IC/CMB model for the inner jet of the PKS 0637$-$752. Despite extensive study of this source since its discovery as an X-ray jet in 1999,  no quantitative model has been published for the inner jet region extending from the 3.4 to 7 arcsec from the core.

The inner jet is modeled as a homogeneous cylinder with diameter $0 \farcs 24$ and length $3 \farcs 6$. We obtain the intrinsic jet diameter from the 20.2 GHz ATCA map by subtracting the full width half maximum of the restoring beam from the observed transverse width in quadrature ($\theta_{\rm true} = \sqrt{\theta_{\rm obs}^2 - \theta_{\rm beam}^2}$). The emission region moves with bulk Lorentz factor $\Gamma$ at angle $\theta$ to the line of sight. The relativistic plasma contains a uniform magnetic field $B$ and an isotropic power-law electron distribution $N(\gamma)\,d\gamma=\kappa\,\gamma^{-m}d\gamma$ between minimum and maximum Lorentz factor ($\gamma_{\rm min}, \gamma_{\rm max}$). The measured radio spectral index $\alpha = 0.56$ (Section~2.3) sets the electron power-law index to $m=2\alpha+1=2.12$. We assume Thomson-regime scattering of cosmic microwave background photons with energy density $\rho_{\rm CMB} = 4 \times 10^{-13} (1+z)^4 \rm erg~ cm^{-3}$.

For each trial viewing angle $\theta$, we computed two independent magnetic field estimates. First, we calculate the minimum-energy field $B_{\rm min}$ that minimizes the total (particle and magnetic) energy densities from the monochromatic radio luminosity at 4.8 GHz and $\alpha=0.56$ following \citet{WorrallBirkinshaw2006}. We assume a unit volume filling factor and equal energy partition between baryons and electrons. Second, we determine the field $B_{\rm cmb}$ required to reproduce the observed radio-to-X-ray flux ratio by inverse Compton scattering using \citet{FeltenMorrison1966}. We iteratively solve for $\Gamma$ such that $B_{\min}=B_{\rm cmb}$ in the jet’s comoving frame. The complete mathematical derivation, including synchrotron emissivity calculations, relativistic transformations, and energy density computations, is provided in Appendix \ref{app:iccmb}.

Next, we follow the Bayesian scheme of \citet{Maithil2025}, drawing $\theta$ from the beaming-aware prior given by \citet{Murphy1988}
\begin{equation}
    f(\theta)\,\mathrm{d}\theta =
\frac{2\,(1-\beta^{2})\,\sin\theta}{\left(1-\beta\cos\theta\right)^{3}}\,
\mathrm{d}\theta ,
\label{Murphy}
\end{equation}
where $\beta$, is the bulk speed of the jet. This prior combines the geometric probability ($\propto\sin\theta$) with the flux-limited selection bias ($\propto\delta^{3}$, where $\delta=[\Gamma(1-\beta\cos\theta)]^{-1}$ is the Doppler factor). Substituting \(\beta\) with \(\sqrt{1-\Gamma^{-2}}\) in equation (\ref{Murphy}) and integrating it over $\theta$ yields the cumulative distribution function \(P(\theta)\) as a function of \(\Gamma\). Because our IC/CMB constraint provides \(\Gamma\) for every trial angle, \(P(\theta)\) is fully specified once \(\theta\) is chosen. 

We draw a uniform deviate $u\in(0,1)$ and solve $P(\theta;\Gamma(\theta)) = u$ to obtain a self–consistent viewing angle. Repeating this procedure 10,000 times generates the posterior distribution of $\theta$. The medians and 90\,\% confidence intervals reported in Section~\ref{sec:results} follow directly from these Monte-Carlo samples.

\section{Results}\label{sec:results}

We adopt a low-energy electron cutoff of $\gamma_{\rm min}$ = 10 and investigate three scenarios distinguished by the high-energy electron cutoff. Case I assumes $\gamma_{\rm max} = 2.5 \times 10^4$ and fits radio and X-ray emission without violating the Fermi upper limits, but the low electron cutoff requires an independent population to produce the synchrotron emission observed by ALMA (as shown in Fig. \ref{fig:SEDpowerlaw}). Case II fits the radio, ALMA and X-ray observed fluxes with $\gamma_{\rm max} = 5 \times 10^4$ but exceeds the Fermi upper limits . Finally, Case III presents the limiting IC/CMB model ($\gamma_{\rm max} = 5 \times 10^4$), which fits the radio and ALMA data while attributing only 22\% of the observed X-ray emission to IC/CMB, thus remaining consistent with the Fermi limits. These scenarios test whether a single electron population can explain the broadband spectrum or whether multiple electron population operates in the inner jet. 

\begin{figure*}[h!]
\gridline{
\includegraphics[width=0.33\columnwidth,trim=0cm 0.0cm 0.cm 0.cm,clip]{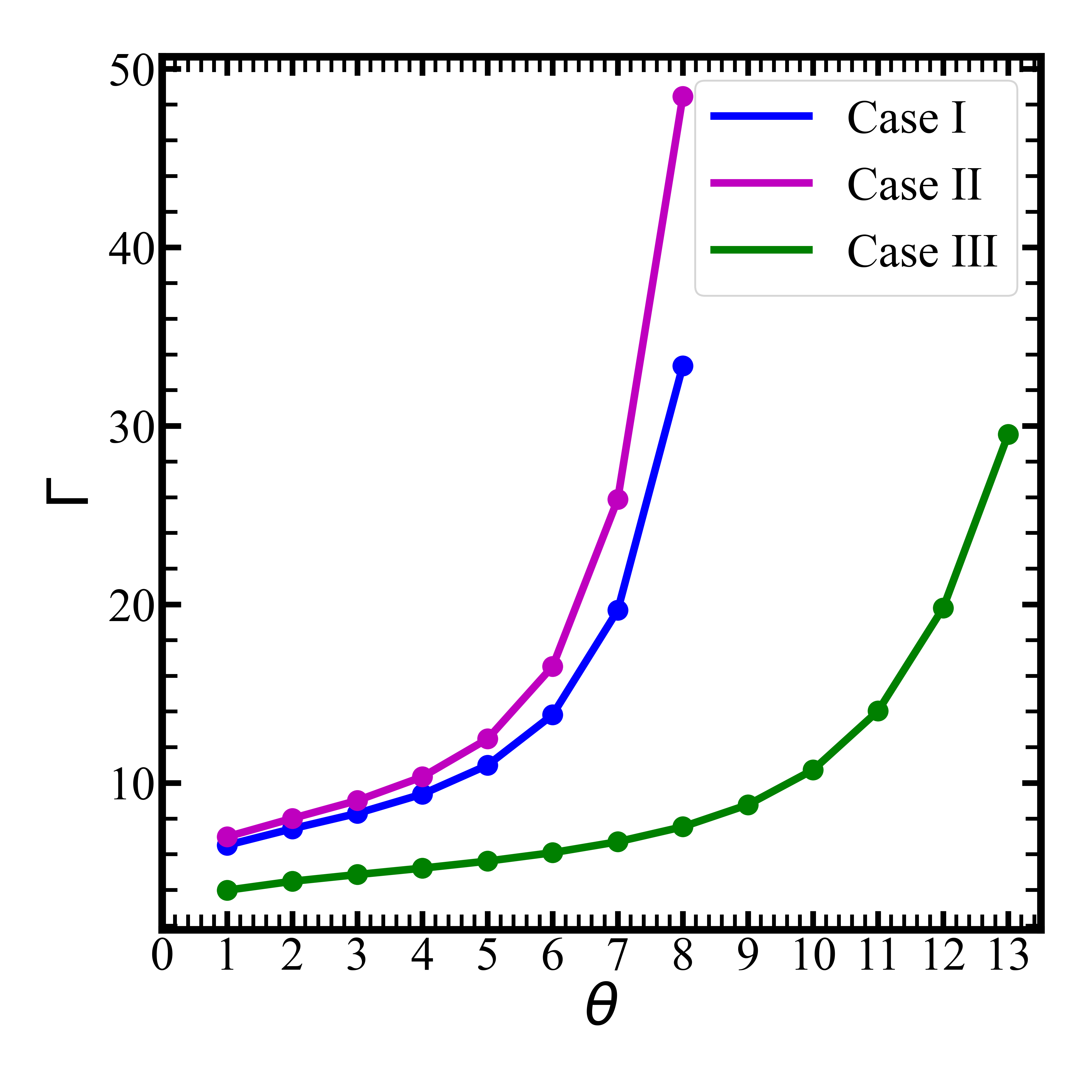}
\includegraphics[width=0.33\columnwidth,trim=0.cm 0cm 0.0cm 0.cm,clip]{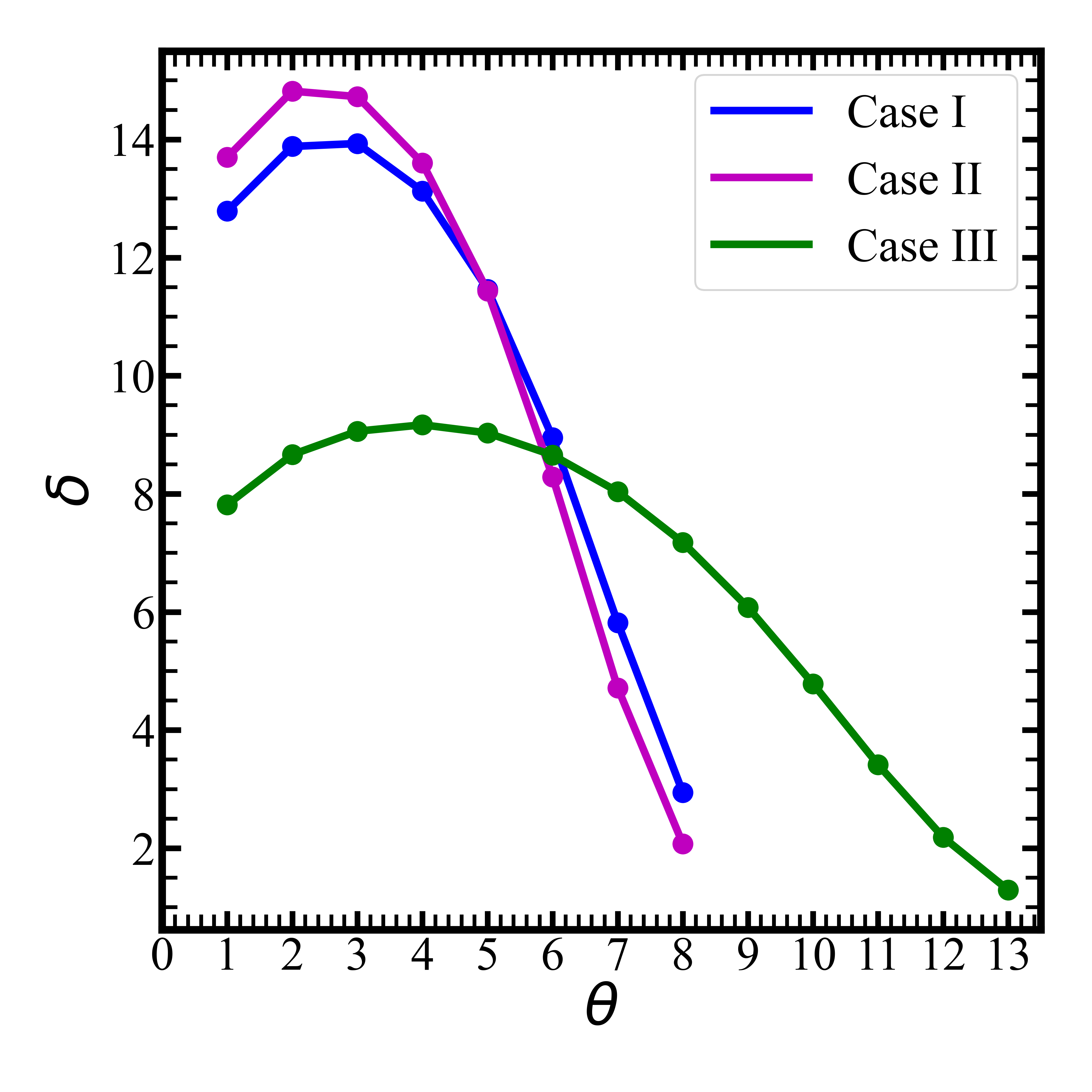}
\includegraphics[width=0.33\columnwidth,trim=0.cm 0.cm 0.0cm 0.cm,clip]{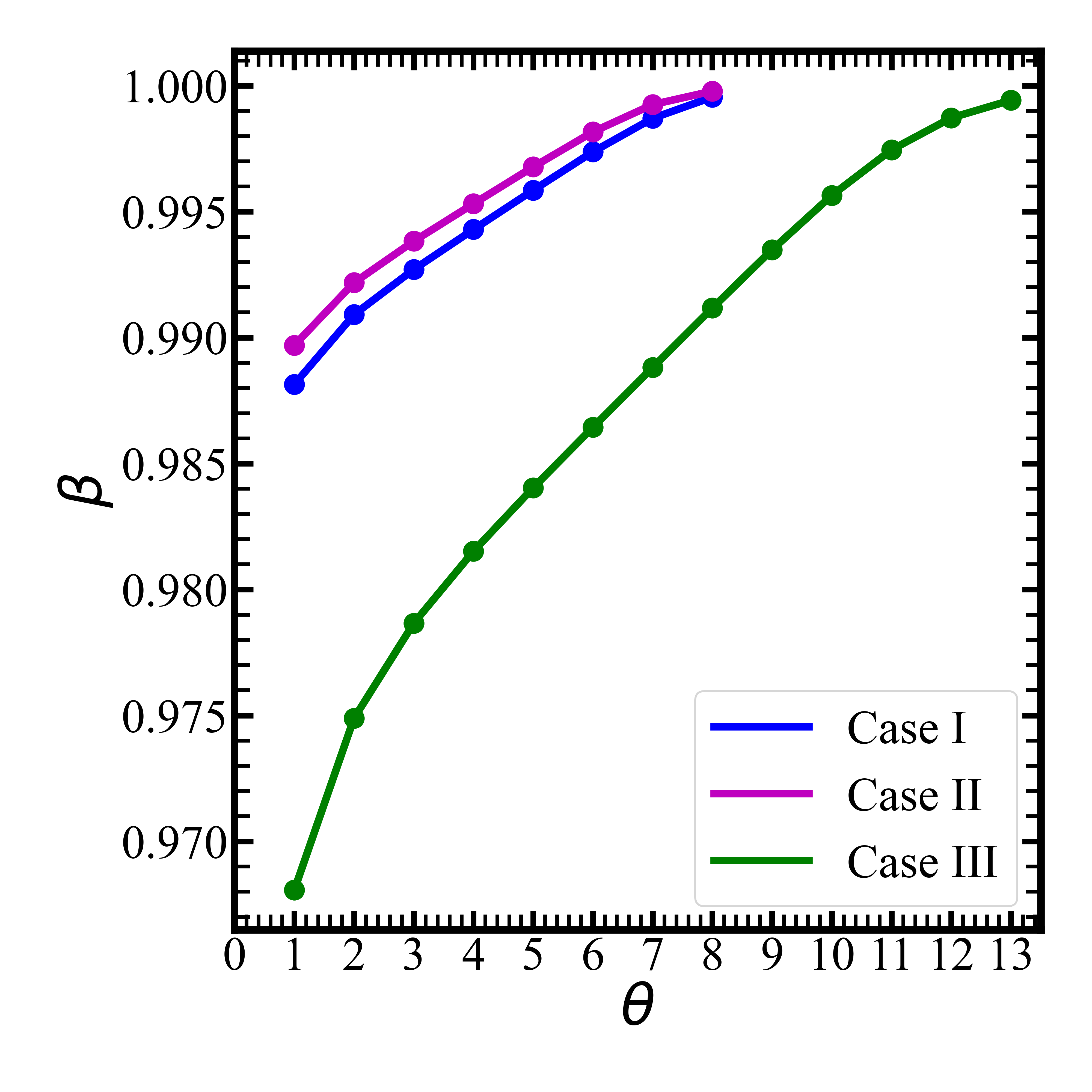}}

\gridline{
\includegraphics[width=0.33\columnwidth,trim=0cm 0.0cm 0cm 0cm,clip]{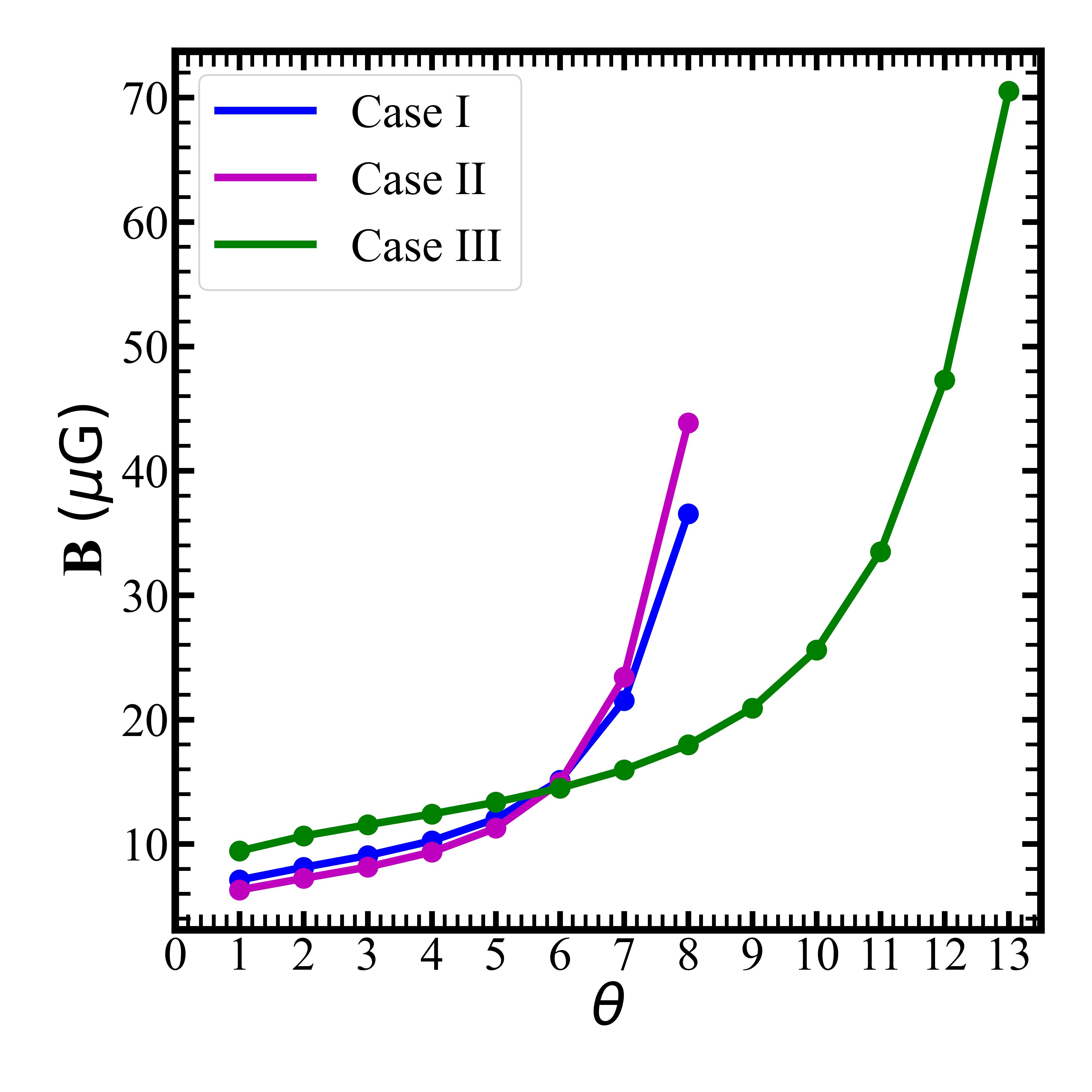}
\includegraphics[width=0.33\columnwidth,trim=0.cm 0.0cm 0cm 0cm,clip]{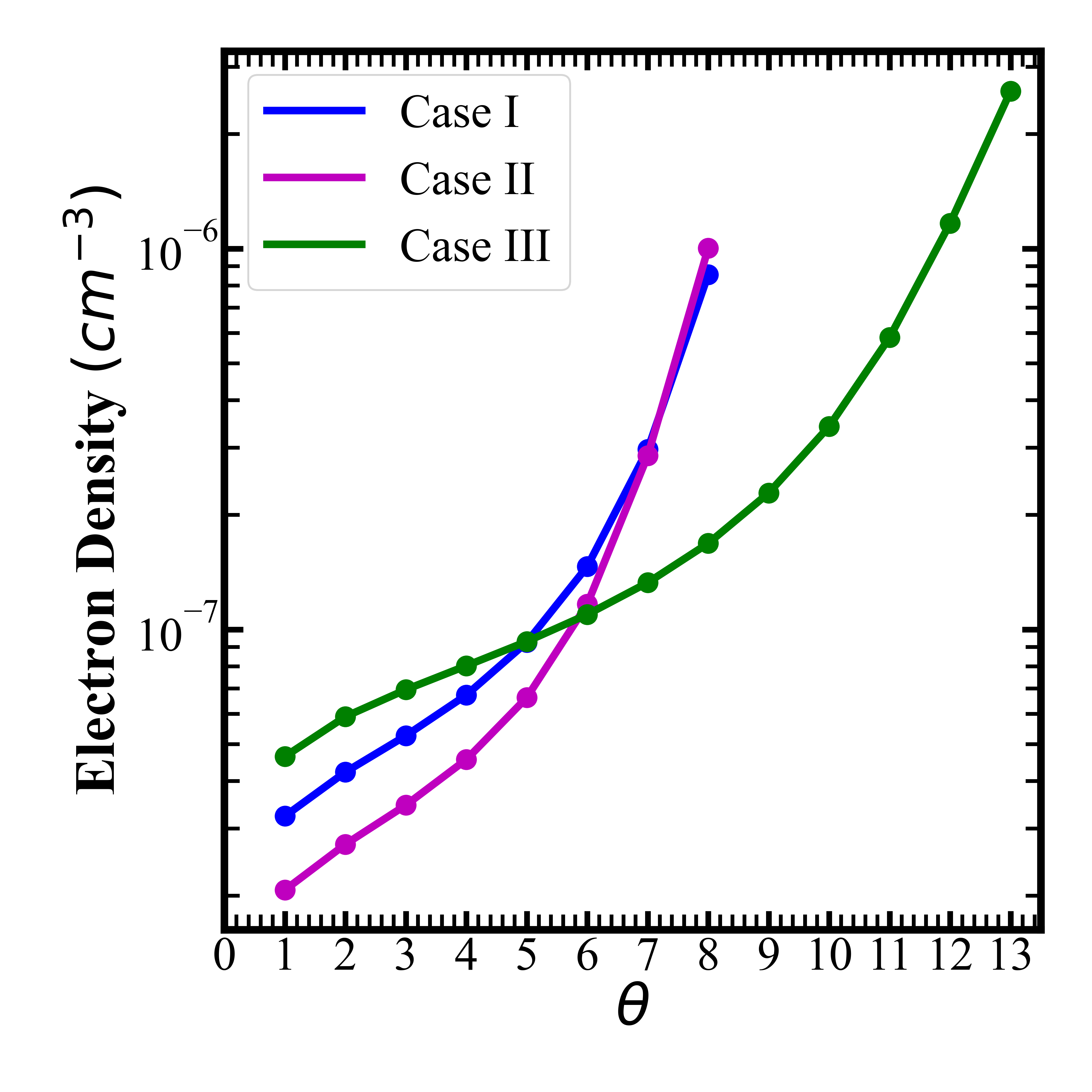}
\includegraphics[width=0.33\columnwidth,trim=0cm 0.0cm 0cm 0cm,clip]{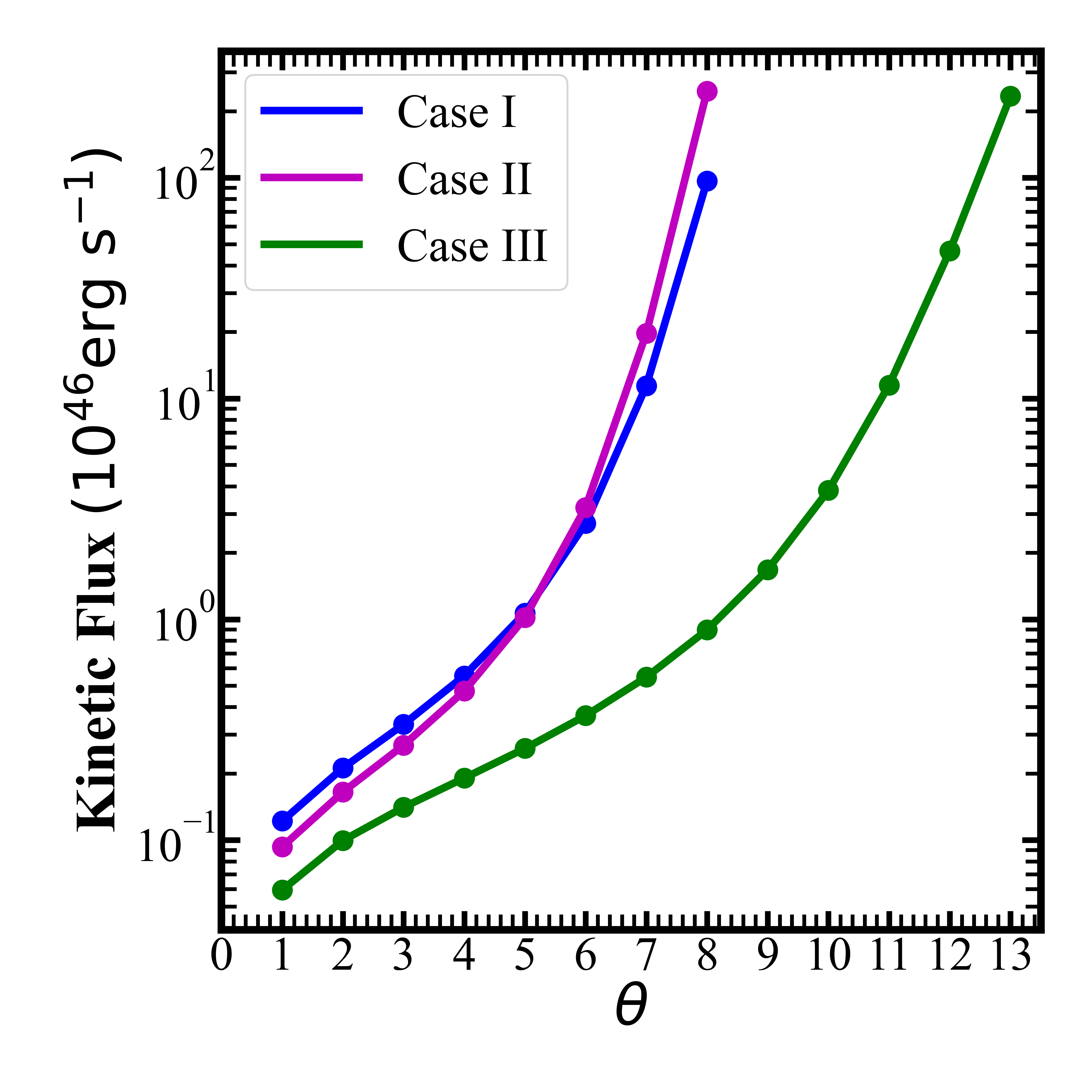}}

\gridline{
\includegraphics[width=0.33\columnwidth,trim=0cm 0.0cm 0cm 0cm,clip]{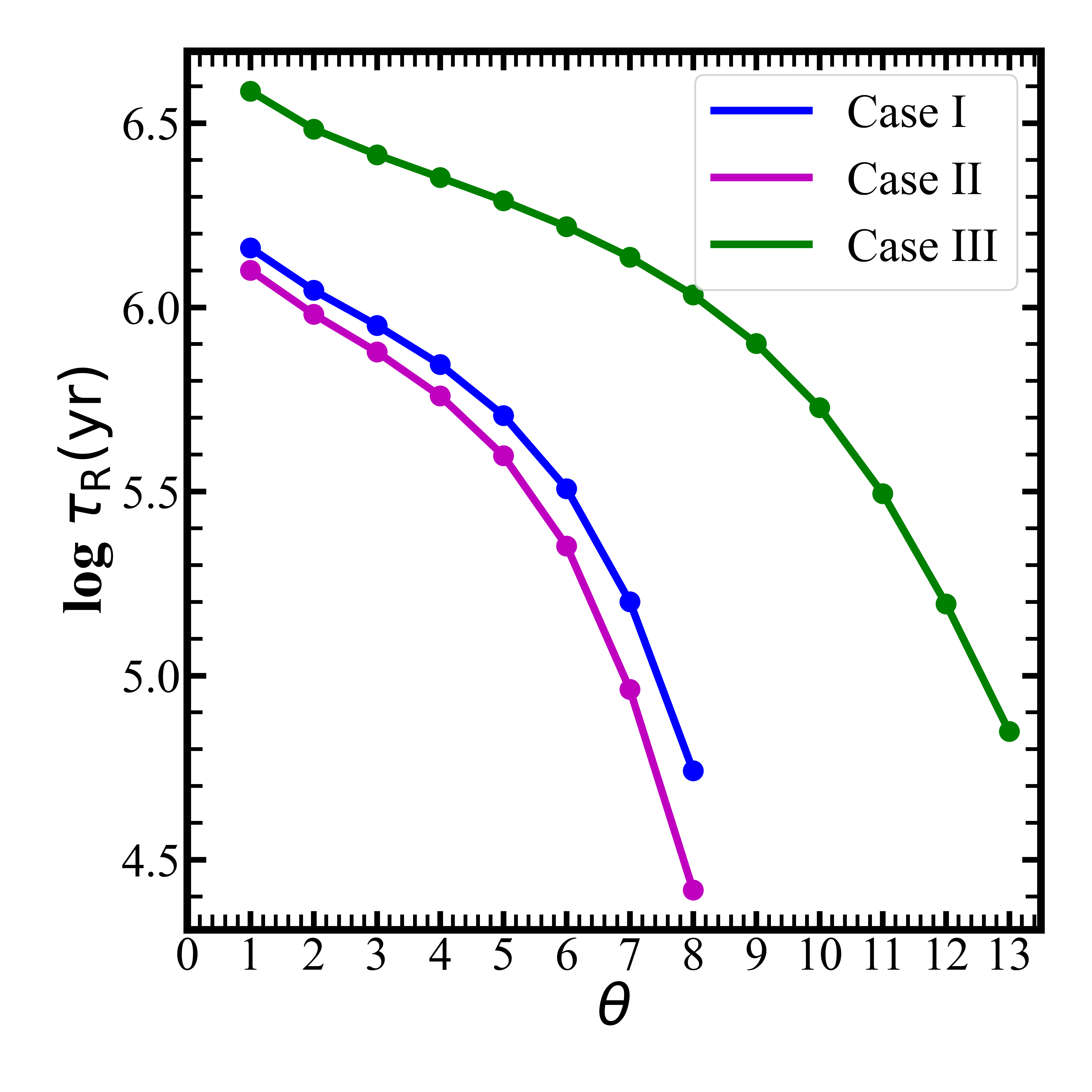}
\includegraphics[width=0.33\columnwidth,trim=0.cm 0.0cm 0cm 0cm,clip]{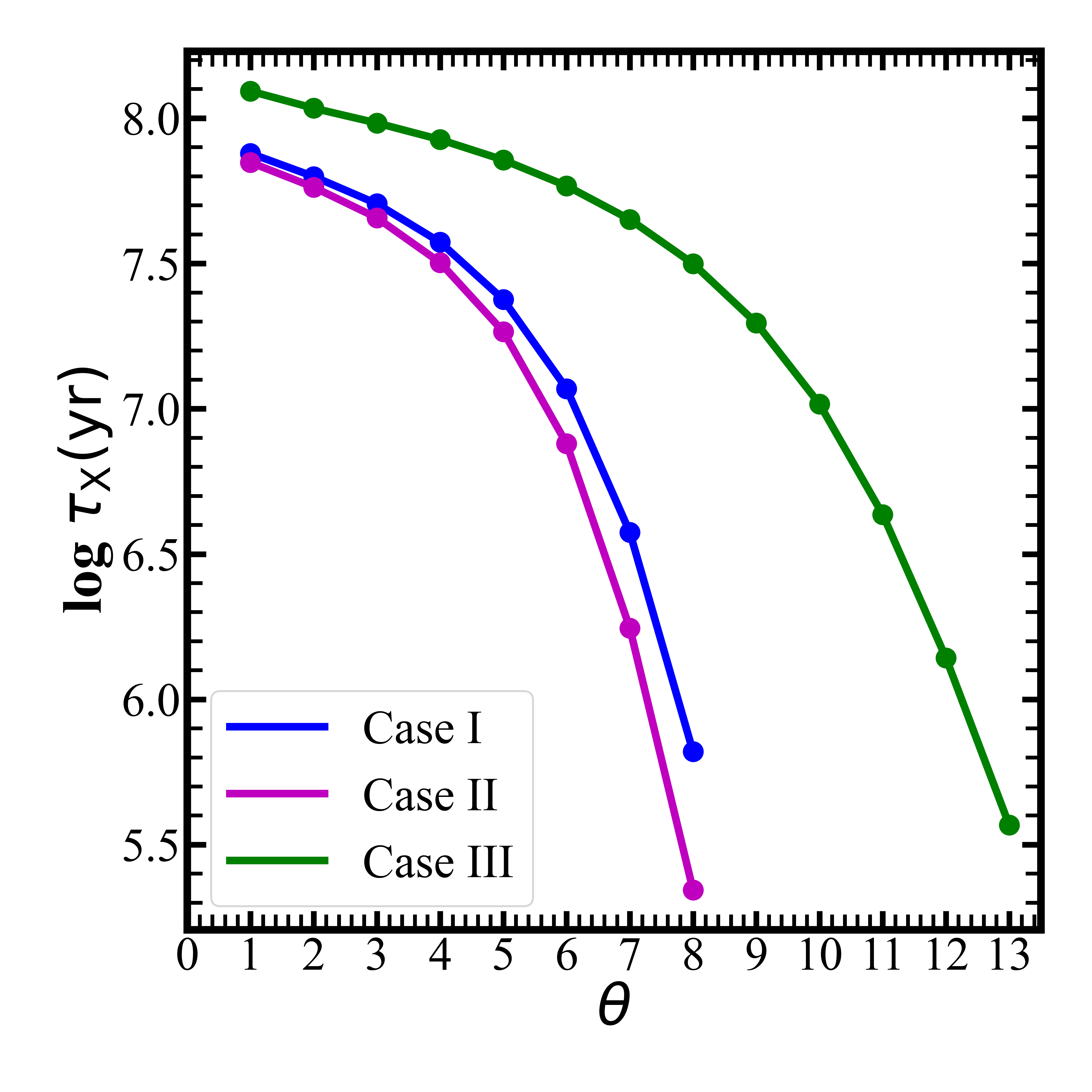}
\includegraphics[width=0.33\columnwidth,trim=0cm 0.0cm 0cm 0cm,clip]{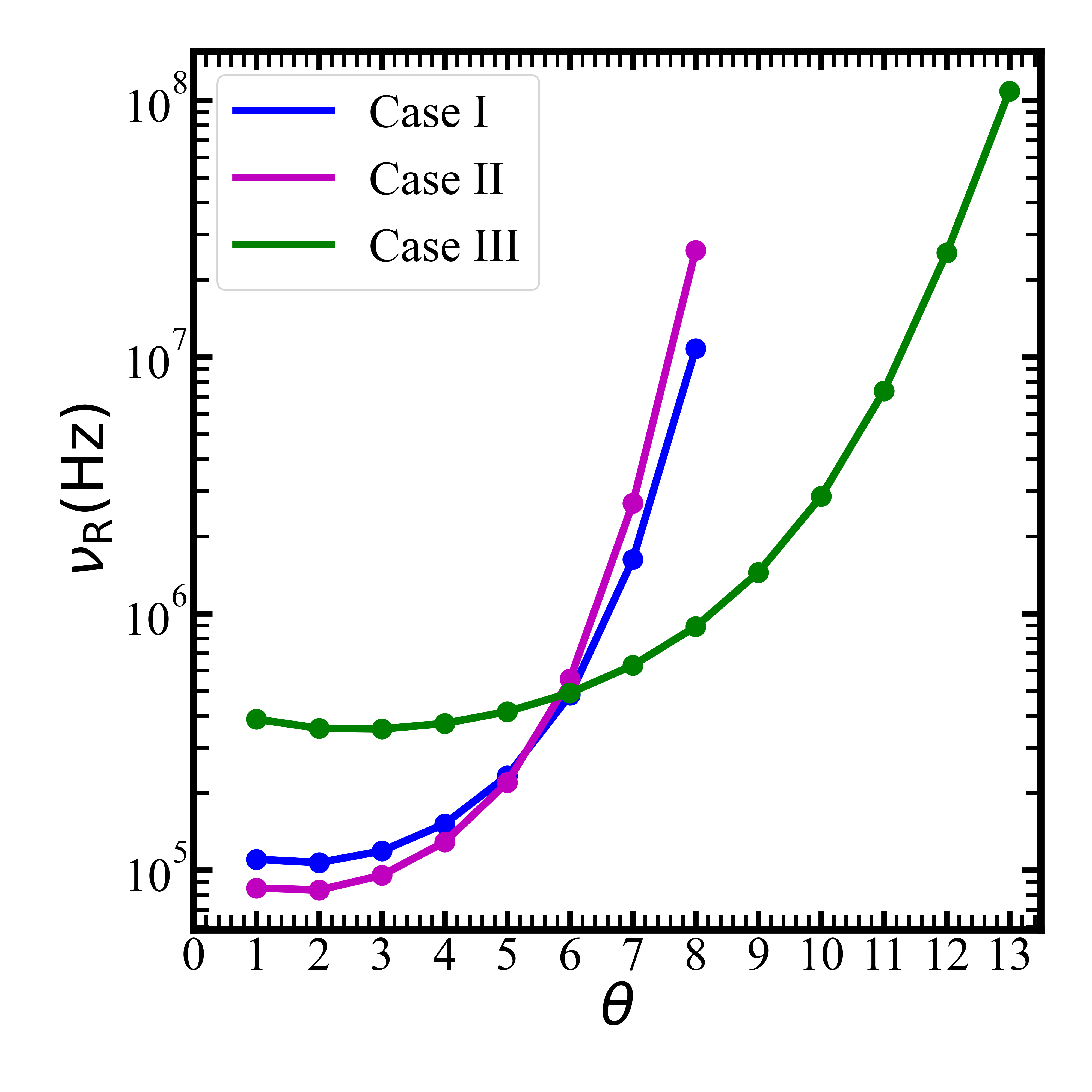}}
\caption{\small IC/CMB model results for PKS 0637$-$752. \textbf{Top:} The relativistic parameters $\Gamma$ (bulk Lorentz factor), $\delta$ (Doppler factor), and $\beta$ (bulk velocity in units of the speed of light) as functions of the viewing angle $\theta$. \textbf{Middle:} The jet parameters, including the magnetic field strength (B), the number density of synchrotron-emitting electrons, and the kinetic energy of the jet, plotted as functions of $\theta$. \textbf{Bottom:} Lifetime of radio-emitting electrons at $\gamma_{\rm max}=5000$, X-ray-emitting electrons via inverse-Compton scattering of CMB at 1 keV, and the radio frequency at which the 1 keV emitting electrons radiate via synchrotron as a function of $\theta$.}
\label{ICCMB}
\end{figure*}

Fig.~\ref{ICCMB} illustrates the results from our IC/CMB model for the three cases. Over the overlapping range $1\degr \leq \theta \leq 8\degr$, the three solutions show systematic differences. The top panel of Fig.~\ref{ICCMB} shows how the bulk Lorentz factor $\Gamma$, Doppler factor $\delta$, and dimensionless speed $\beta\equiv v/c$ vary with line-of-sight angle $\theta$ for the three cases. $\Gamma$ rises from 6 to 33 (Case~I) and $7$ to 48 (Case~II), whereas Case~III remains modest, $4$ to 7 (continuing to 29 by $13\degr$). At fixed $\theta\!\le\!8\degr$, $\Gamma_{\rm Case~ II}>\Gamma_{\rm Case~ I}\gg\Gamma_{\rm Case~ III}$. $\delta$ peaks at $\theta\!\approx$3$\degr$ for Case~I and at $\theta\!\approx$2$\degr$ for Case~II with values of $\simeq14$ and $\simeq15$, respectively. For Case~III $\delta \simeq9$ at $\theta\!\approx\!4\degr$. By $8\degr$ $\delta$ falls to $2$ and 3 for Case~I and Case~II, receptively, but remains $7$ in Case~III, then declines to $1$ by $13\degr$. The jets remain highly relativistic in all three cases. For Cases I and II, $\beta$ spans $\approx$0.990$–0.999$ across $1\degr–8\degr$. Case III varies more gradually, from $\beta \approx 0.970$ to 0.999 over $1\degr–13\degr$.

The middle panel of Fig.~\ref{ICCMB} shows how the intrinsic jet properties predicted by our IC/CMB model as a function of viewing angle. Magnetic field $B$ ($\mu$G) within $1$–$8\degr$, $B$ grows from 7.1 to 36.5 (Case~I) and 6.3 to 43.8 (Case~II); Case~III starts higher at small angles, 9.4 at $1\degr$, but reaches only $18.0$ at $8\degr$ before increasing rapidly to 70.5 by $13\degr$. Electron density $n_e (\rm cm^{-3})$ rises from 3.2$\times10^{-8}$ to 8.5$\times10^{-7}$ (Case~I) and 2.1$\times10^{-8}$ to 1.0$\times10^{-6}$ (Case~II) over $1$–$8\degr$; Case~III is highest at $1\degr$ (4.6$\times10^{-8}$) but remains lower at $8\degr$ ($1.7\times10^{-7}$), then climbs to 2.6$\times10^{-6}$ at $13\degr$. Kinetic‐energy budget (in units of $\rm erg~ s^{-1}$): Case~II is the most demanding over $1$–$8\degr$ (9.3 $\times 10^{44}$ - 2.5$\times 10^{48}$), exceeding Case~I (1.2 $\times 10^{45}$ - 9.7$\times 10^{47}$), while Case~III stays low in that interval (5.9 $\times 10^{44}$ - 8.9$\times 10^{45}$) but grows steeply at larger angles, reaching 2.3$\times 10^{48}$ by $13\degr$.

In summary, Cases~I and II share similar beaming (peaks near $\theta\!\sim\!2\degr-3\degr$), with Case~II requiring systematically larger $\Gamma$, $B$, $n_e$, and kinetic power—consistent with its tension with the \textit{Fermi} limits. Case~III achieves the observed radio/ALMA emission with substantially lower beaming ($\delta$) and energy requirements for $\theta\!\le\!8\degr$, while allowing a broader orientation range ($\le\!13\degr$); the associated increase in $B$, $n_e$, and kinetic power at larger $\theta$ compensates for reduced Doppler boosting.

Using Equation (4) of \citet{Worrall2020}, which applies when inverse Compton losses outweigh synchrotron losses, we estimate the relevant electron timescales and frequencies for the inner jet. The bottom panel of Fig.~\ref{ICCMB} indicates that synchrotron emitting electrons with $\gamma = 5000$ have $\tau_R \simeq 1 \times 10^{6} \text{--} 5 \times 10^{4}\,\mathrm{yr}$ for Case~I and $\tau_R \simeq 1 \times 10^{6} \text{--} 3 \times 10^{4}\,\mathrm{yr}$ for Case~II over $1\degr \leq \theta \leq 8\degr$. For Case~III, $\tau_R \simeq 4 \times 10^{6} \text{--} 7 \times 10^{4}\,\mathrm{yr}$  over $1\degr \leq \theta \leq 13\degr$. In contrast, electrons that up-scatter CMB photons to $1\,\mathrm{keV}$ survive much longer, with lifetimes of $\tau_{X} \approx 8 \times 10^{7}\text{--} 7 \times 10^{5}\,\mathrm{yr}$ and $\tau_{X} \approx 7 \times 10^{7}\text{--} 2 \times 10^{5}\,\mathrm{yr}$ for Case~I and Case~II, respectively, across the same angular range of $1\degr \leq \theta \leq 8\degr$. For Case~III, $\tau_{X} \approx 1 \times 10^{8}\text{--} 4 \times 10^{5}\,\mathrm{yr}$ over $1\degr \leq \theta \leq 13\degr$. This confirms that the IC/CMB emitting population ages more slowly than its radio counterpart. The same electrons would emit synchrotron radio at $\nu_{\mathrm{R}} \simeq 1 \times 10^5 \text{--} 1 \times 10^7$ Hz and $\nu_{\mathrm{R}} \simeq 8 \times 10^4 \text{--} 3 \times 10^7$ Hz for Case~I and Case~II, respectively over $1\degr \leq \theta \leq 8\degr$. Whereas, for Case~III, $\nu_{\mathrm{R}} \simeq 4 \times 10^5 \text{--} 1 \times 10^8$ Hz over $1\degr \leq \theta \leq 13\degr$.

\begin{figure*}[h!]
\gridline{
\includegraphics[width=0.33\columnwidth,trim=0cm 0.0cm 0.cm 0.cm,clip]{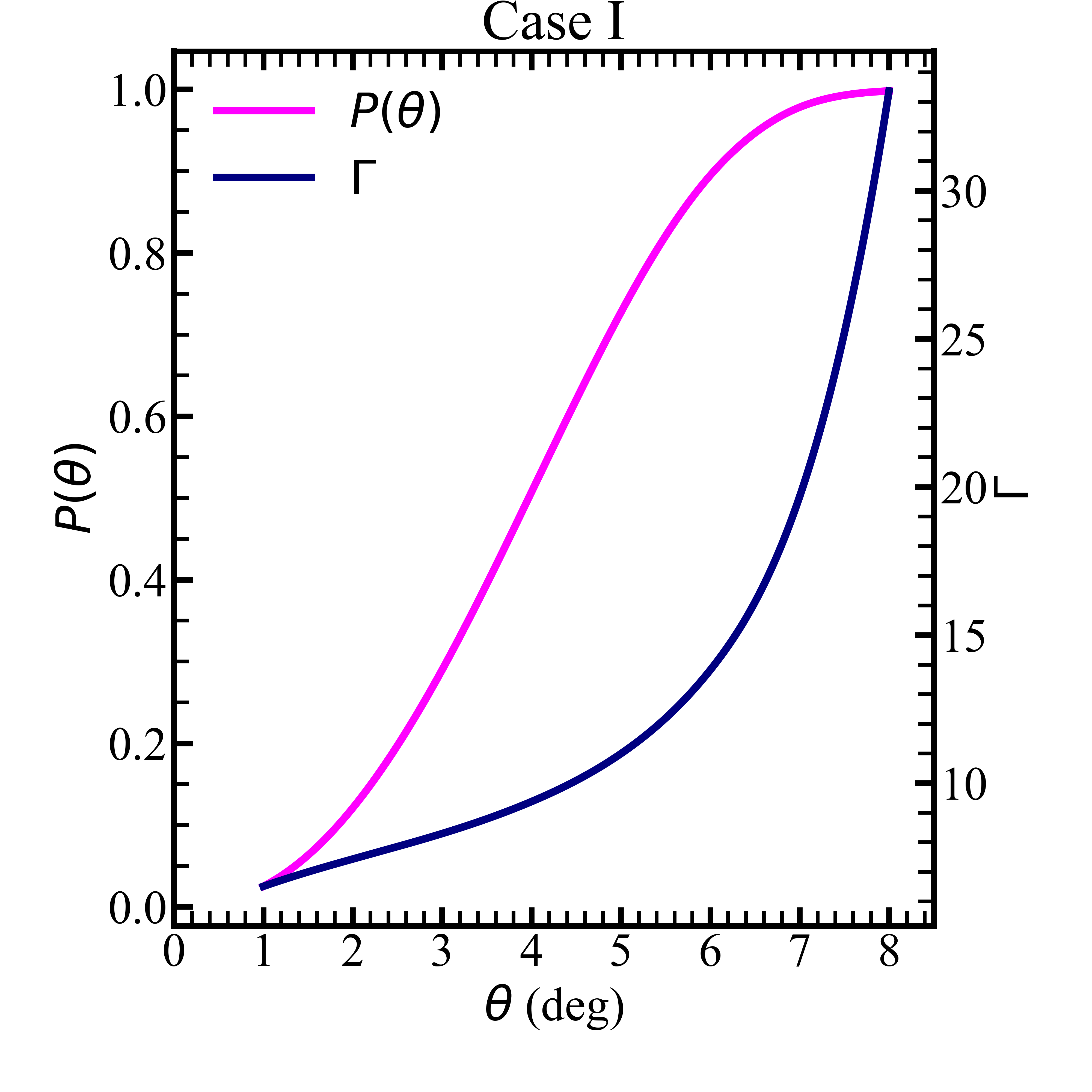}
\includegraphics[width=0.33\columnwidth,trim=0.cm 0cm 0.0cm 0.cm,clip]{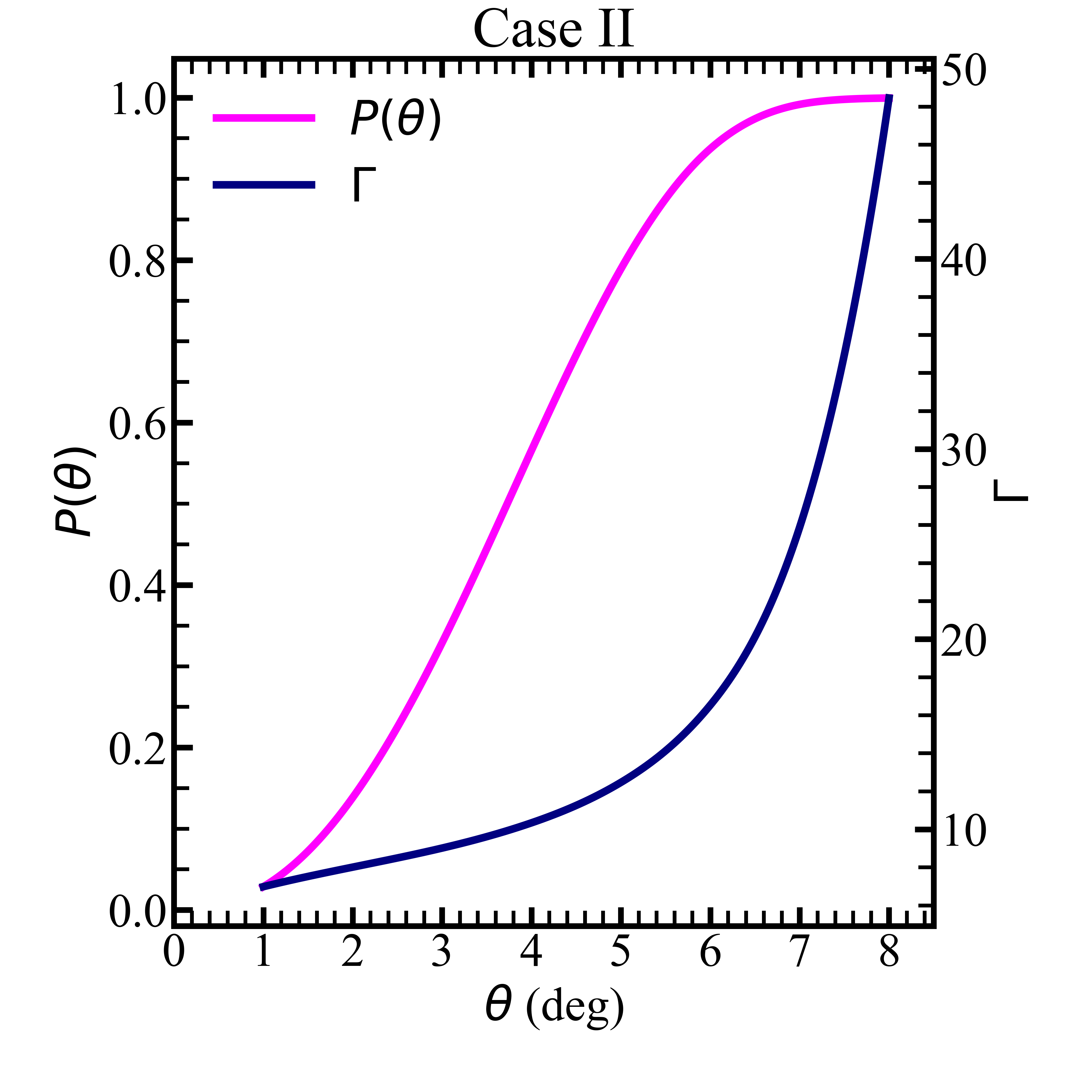}
\includegraphics[width=0.33\columnwidth,trim=0.cm 0.cm 0.0cm 0.cm,clip]{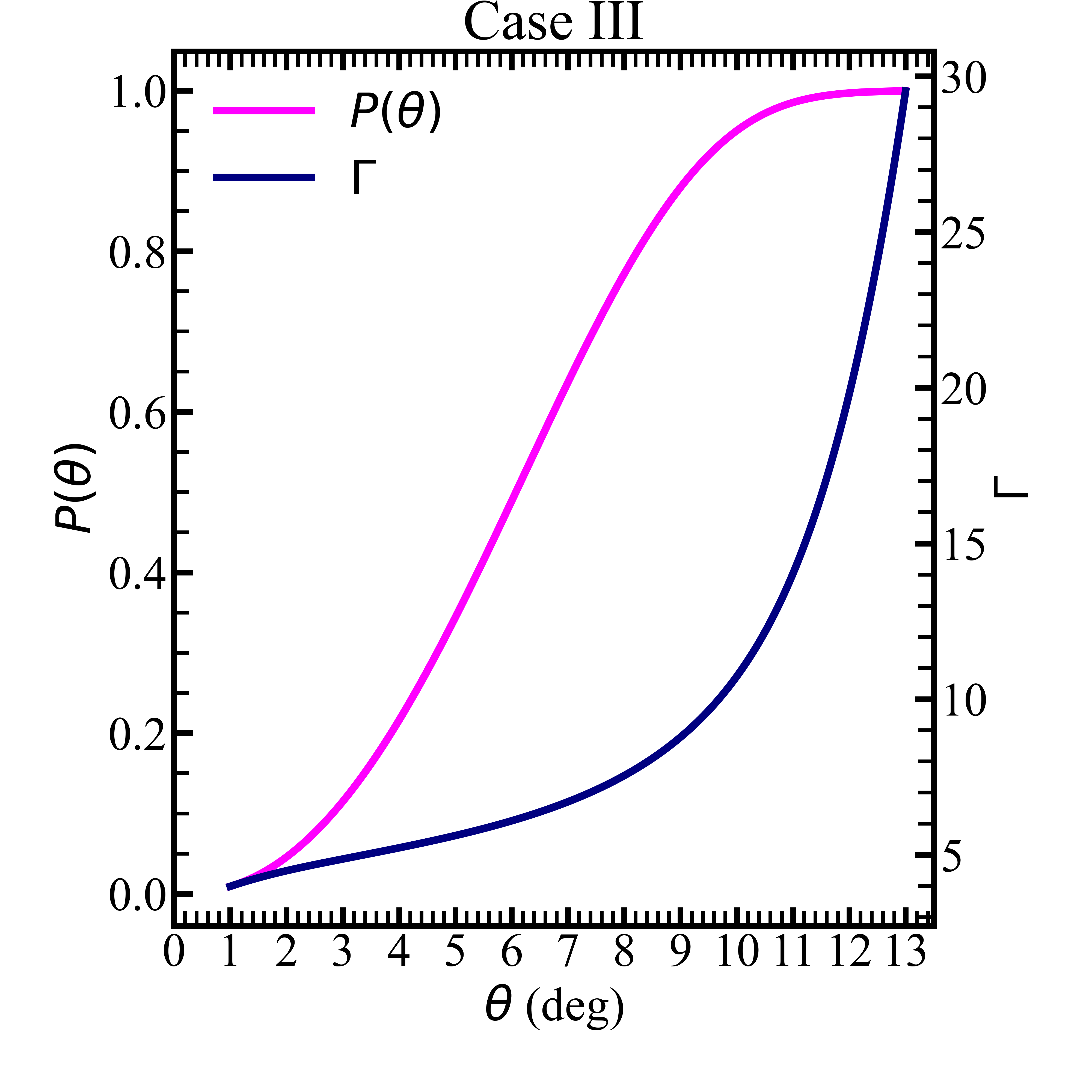}}
\gridline{
\includegraphics[width=0.33\columnwidth,trim=0cm 0.0cm 0cm 0cm,clip]{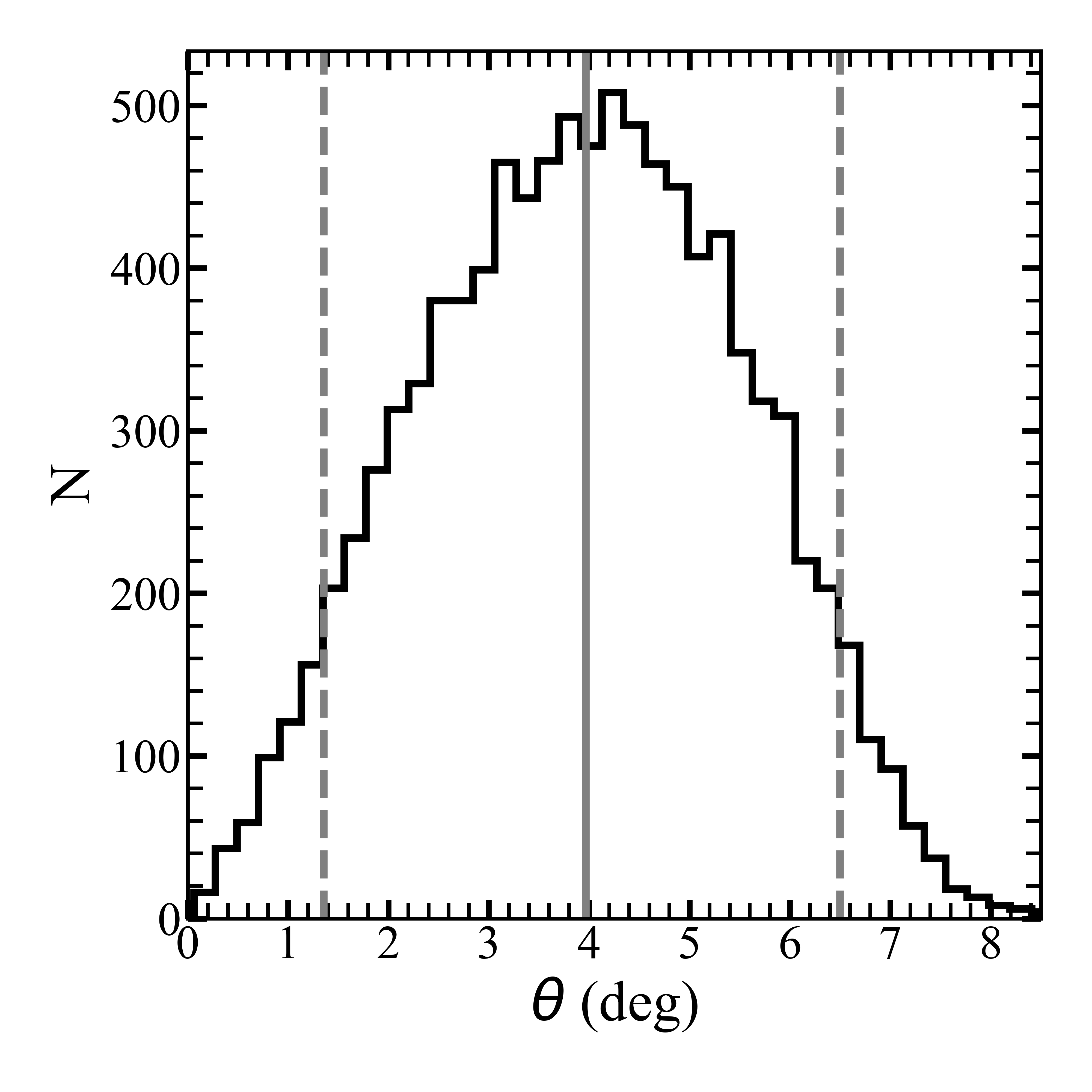}
\includegraphics[width=0.33\columnwidth,trim=0.cm 0.0cm 0cm 0cm,clip]{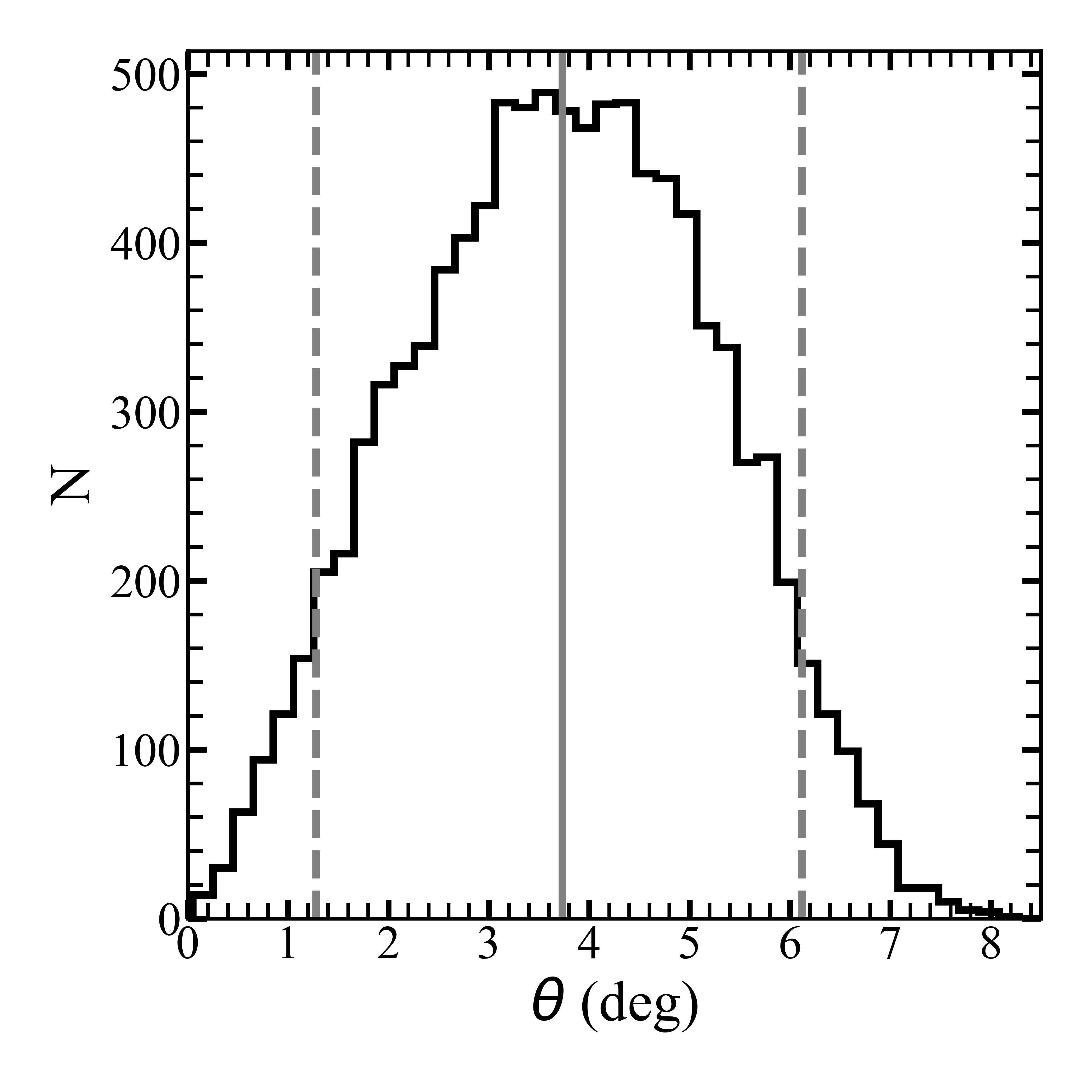}
\includegraphics[width=0.33\columnwidth,trim=0cm 0.0cm 0cm 0cm,clip]{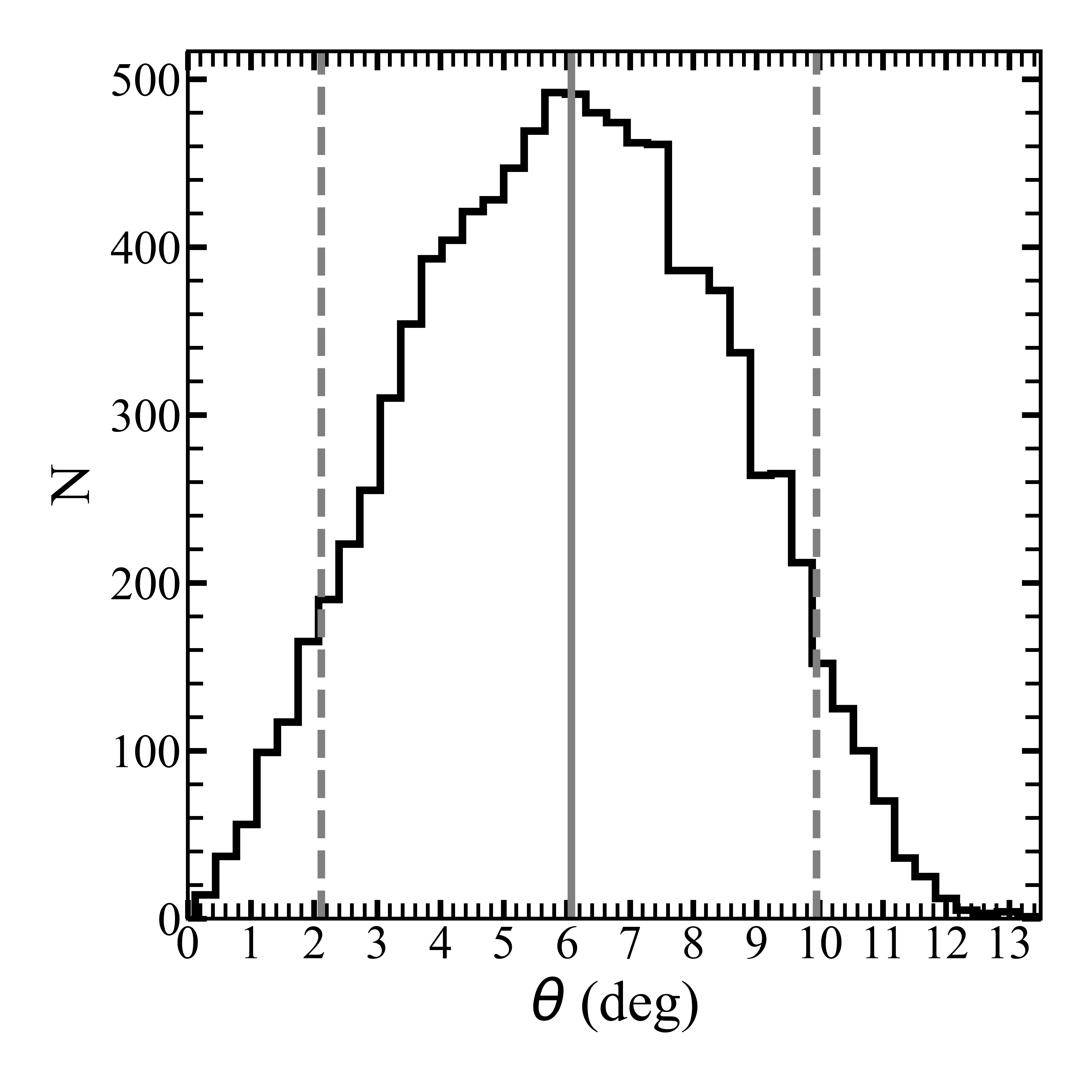}}
\caption{\small Top: Probability (magenta) and bulk Lorentz factor (blue) as a function of line of sight angle. Bottom: Distribution of line of sight angle $\theta$ based on our Bayesian method. Solid grey lines represents the median $\theta$ and dot-dash lines represent the 90\% confidence intervals.}
\label{Prob}
\end{figure*}

We convert the angle-dependent Lorentz factor predicted by the IC/CMB model into a cumulative distribution function,$P(\theta)$, following the procedure in Section~3. In the upper panel of Fig.~\ref{Prob}, the cumulative probability rises rapidly, reaching unity by \(\theta\simeq8\degr\) for Cases~I and II and by \(\theta\simeq13\degr\) for Case~III, indicating that larger viewing angles are strongly disfavored. The lower panel of Fig.~\ref{Prob} shows the posterior distribution of \(\theta\) from our Monte Carlo sampling: the solid line marks the median and the dashed lines enclose the 90\% credible interval. For Cases~I and II we obtain a median \(\theta\simeq4\degr\) with a 90\% interval of \(\simeq1\degr\)–\(6\degr\); for Case~III the median is \(\simeq6\degr\) with a 90\% interval of \(\simeq2\degr\)–\(10\degr\). We evaluate model parameters at the posterior-median viewing angle and quote ranges corresponding to the 90\% credible interval; the resulting medians and intervals are summarized in Table~\ref{tab:Medianvalues}. Following \citet{Maithil2025}, we use the Jet SED Modeler and Fitting Tool (\texttt{JetSet}; \citealt{Massaro+2006, Tramacere+2009, Tramacere+2011, Tramacere2020}) to obtain the SEDs with jet parameters fixed to the values derived above.

\begin{deluxetable}{lll}[h!]
\tablecaption{Constraints on jet parameters using $\alpha=0.56, \gamma_{\rm min} = 10$.\label{tab:Medianvalues}} 
\tablehead{
\colhead{Parameters} & \colhead{Median} & \colhead{90\% C.I.}}
\startdata
\cutinhead{\textbf{Case I}}
$\theta$ & 4.0 & (1.3, 6.5) \\
$\Gamma$ &  9.3 & (6.9, 16.1) \\
$\delta$ & 13.2 & (13.3, 7.4) \\
$\beta$	 &  0.994 & (0.989, 0.998) \\
B ($\mu$G)	& 10.2 & (7.5, 17.6) \\
$n_e~ (10^{-8} \rm cm^{-3})$ & 6.7 & (3.6, 19.4)\\
K.E. $(10^{45} \rm erg~ s^{-1})$ & 5.5 & (1.7, 28.9)\\
\cutinhead{\textbf{Case II}}
$\theta$ & 3.7 & (1.3, 6.1) \\
$\Gamma$ &  9.9 & (7.3, 17.3) \\
$\delta$ & 14.0 & (14.1, 7.9) \\
$\beta$	 &  0.995 & (0.990, 0.998) \\
B ($\mu$G)	& 9.0 & (6.6, 15.6) \\
$n_e~ (10^{-8} \rm cm^{-3})$ & 4.2 & (2.3, 12.5)\\
K.E. $(10^{45} \rm erg~ s^{-1})$ & 5.3 & (1.5, 21.6)\\
\cutinhead{\textbf{Case III}}
$\theta$ & 6.1 & (2.1, 9.9)\\
$\Gamma$ &  6.1 & (4.5, 10.6) \\
$\delta$ &  8.6 & (8.7, 4.8) \\
$\beta$	 &  0.987 & (0.975, 0.995) \\
B ($\mu$G)	& 14.6 & (10.8, 25.3) \\
$n_e~ (10^{-8} \rm cm^{-3})$ & 11.1 & (6.0, 33.3) \\
K.E. $(10^{45} \rm erg~ s^{-1})$ &  3.8 & (1.0, 36.0)
\enddata
\end{deluxetable}

\section{Discussion and Conclusions}\label{sec:discussion}

\begin{figure*}[h!]
\gridline{
\includegraphics[width=0.52\columnwidth,trim=0cm 0.0cm 0cm 0cm,clip]{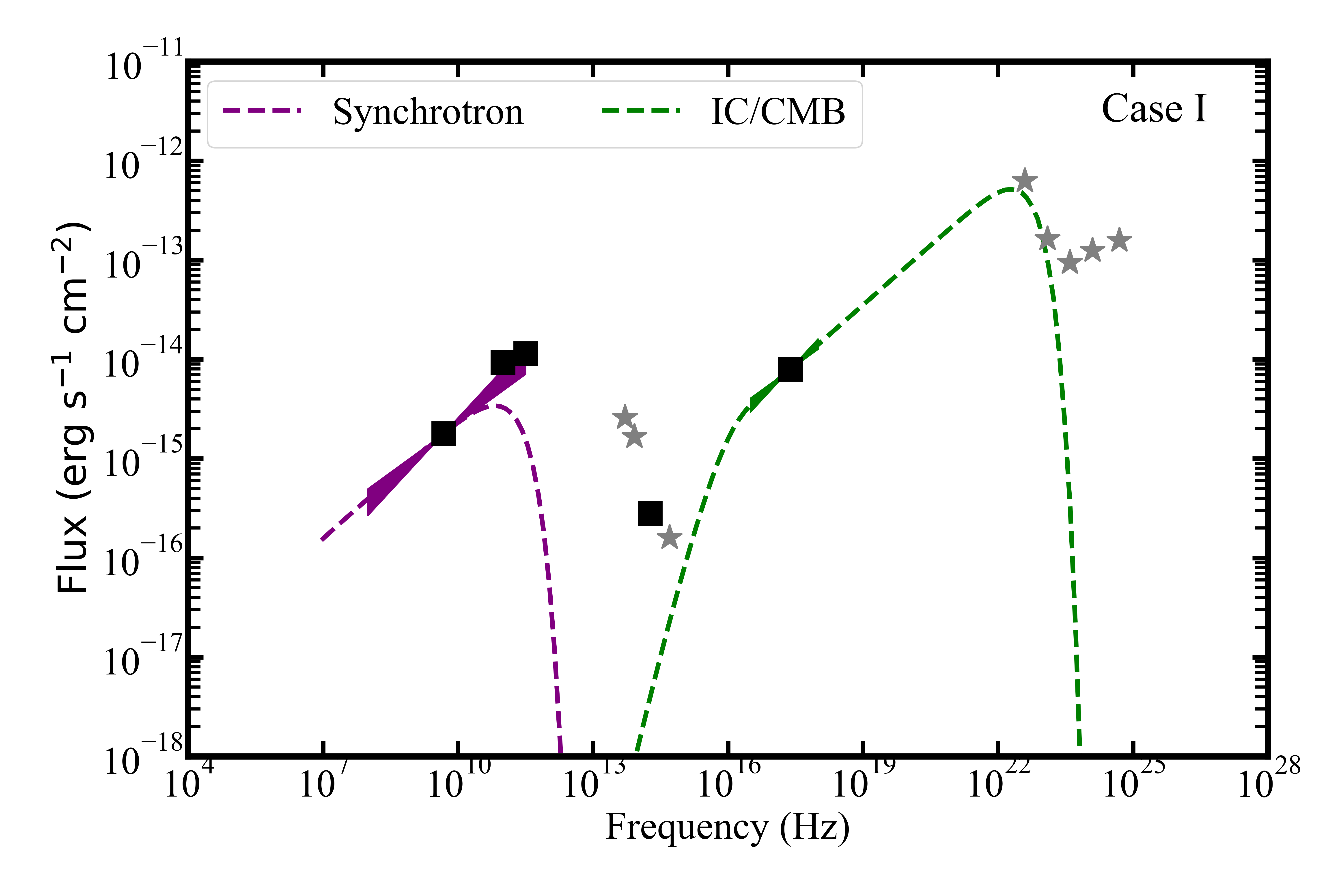}
\includegraphics[width=0.52\columnwidth,trim=0cm 0.0cm 0cm 0cm,clip]{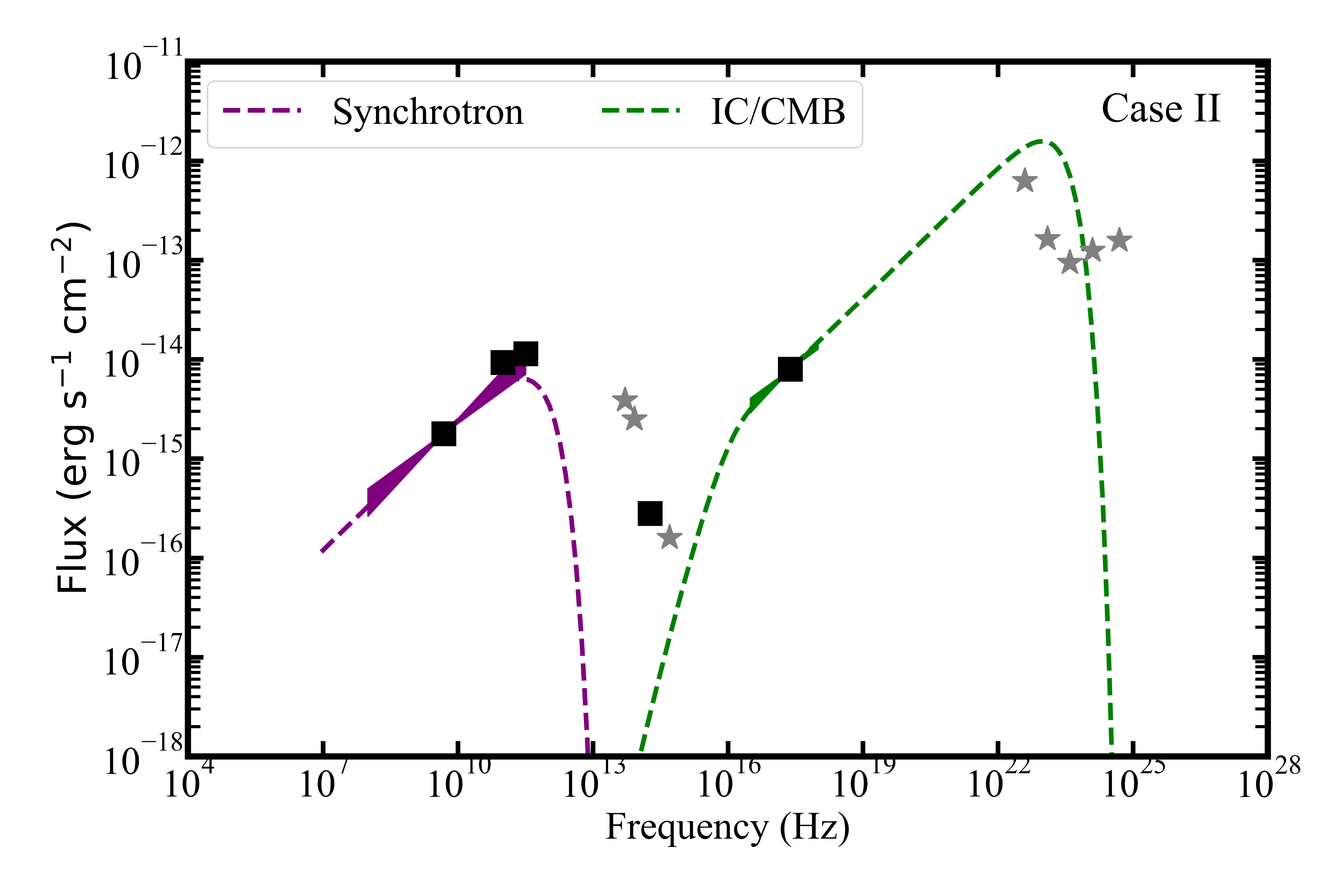}}
\centerline{\includegraphics[width=0.52\columnwidth,trim=0cm 0cm 0cm 0cm,clip]{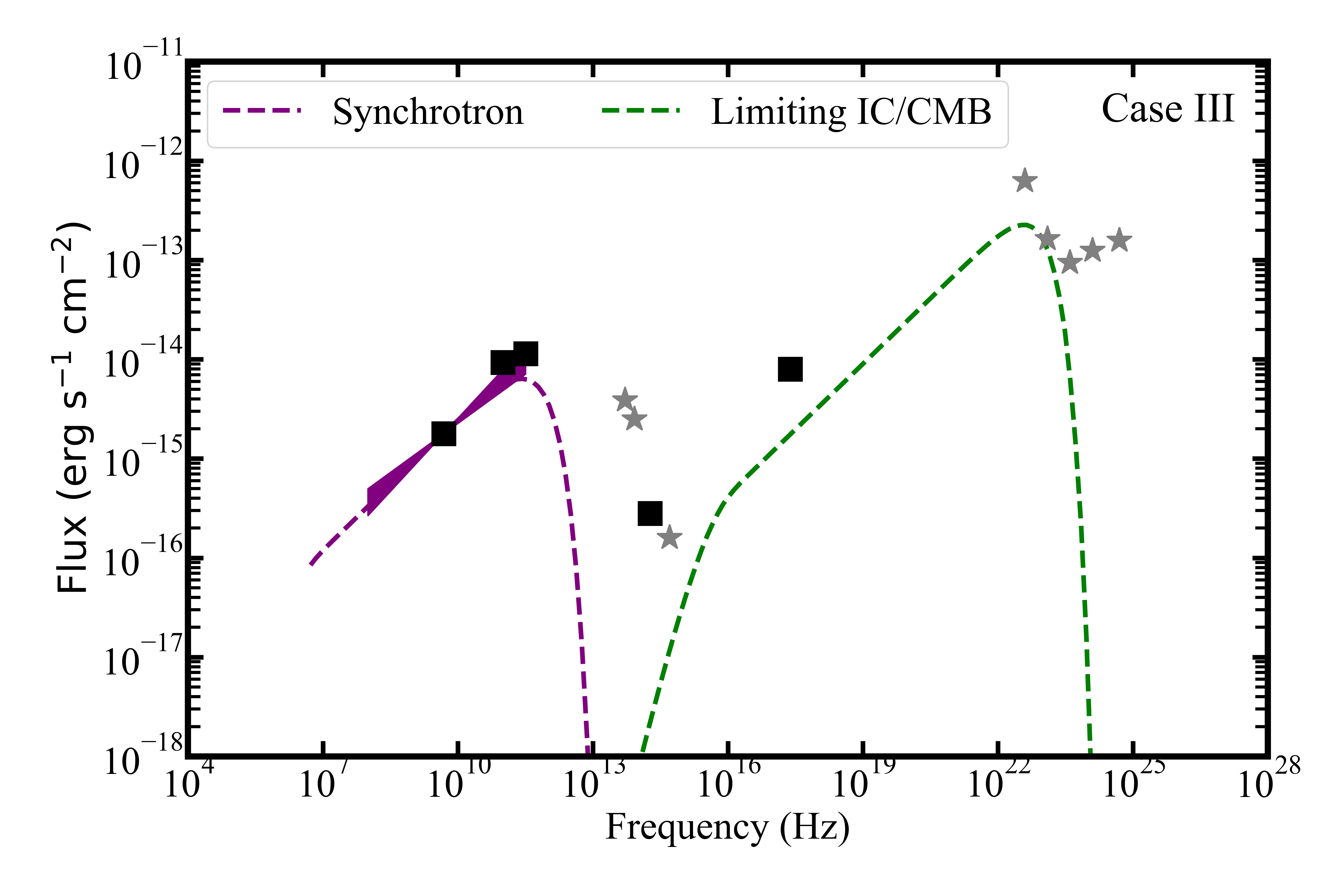}}
\caption{\small SED of inner jet of PKS 0637$-$752 modeled with JetSet using the IC/CMB parameters for the three cases listed in Table 4. Squares show the measured fluxes, and stars indicate upper limits presented in Section 2.} 
\label{fig:SEDpowerlaw}
\end{figure*}

Fig. \ref{fig:SEDpowerlaw} compares the SEDs of the inner jet under three IC/CMB cases that differ primarily in electron high energy cut-off. Case~I ($\gamma_{\rm max}=2.5\times10^{4}$) reproduces the radio and X-ray fluxes without violating the \textit{Fermi}-LAT upper limits, but cannot account for the ALMA-band synchrotron emission without invoking an additional high-energy electron population. Case~II ($\gamma_{\max}=5\times10^{4}$) fits the radio, ALMA, and X-ray data within the uncertainty in the spectral index, but exceeds the $\gamma$-ray flux limits (in three energy bands). Our limiting IC/CMB solution, Case~III (also with $\gamma_{\max}$) fits the radio and ALMA data while assigning only $\simeq 22\%$ of the 1~keV flux to IC/CMB, thus being consistent with Fermi upper limits; the remaining X-rays must arise from a distinct electron population.  The small median viewing angles (Cases~I/II: \(\sim4\degr\); Case~III: \(\sim6\degr\)) imply modest but significant beaming, with Case~II demanding systematically larger \(\Gamma\) and \(\delta\), and therefore the highest kinetic-power budget than in the other cases. This helps explain why, despite fitting the radio–mm–X-ray SED, Case~II exceeds the Fermi upper limits (in three bands). 

These results indicate a hybrid (synchrotron+IC/CMB) origin for the inner-jet X-rays. The relatively flat radio spectrum of the inner jet ($\alpha\!\approx\!0.56$) under predicts the 1~keV flux if modeled as a synchrotron alone, motivating an IC/CMB contribution from a relativistic flow. 
The Fermi upper limits, however, show that a single electron population cannot account for the broadband emission. Case II reproduces ALMA fluxes and exceeds the Fermi limit and is therefore disfavored. In Case~I an additional high-energy population is needed to produce the ALMA-band synchrotron, while in Case~III a distinct population must supply the remainder of the X-rays after the IC/CMB component is curtailed to satisfy the Fermi limits.

Two further points support this case. The X-ray photon index of the inner jet, $\Gamma=1.92~ (\alpha_{\rm X} = 0.92)$, is steeper than the radio spectral index ($\alpha_{\rm R} = 0.56$). For IC/CMB emission from a single unbroken power-law  electron distribution we expect $\alpha_{\rm X}= \alpha_{\rm R}$. In addition, the HST/NICMOS detection of \citet{Mehta2009} at $1.87 \times 10^{14}$ Hz lies well above the synchrotron cutoff of all three models and cannot be produced by the same radio-emitting electron population.  A full two-zone model for this second component is beyond the scope of this work. 

This interpretation is consistent with the spine-sheath flow inferred by \citet{Mehta2009}. From the faint infrared emission of the combined WK 5.7 \& WK 6.3 feature, located within the inner jet, they constrained the power carried by leptons that are cold in the comoving frame, which scatter CMB photons solely through the bulk motion of the jet. For $\Gamma \gtrsim 10$ their upper limits become insensitive to bulk Lorentz factor at a given viewing angle and are set by $\theta$, ranging from $\sim 10^{45}~ \rm erg~ s^{-1}$ at $\theta =4\degr$ to $\sim 10^{47}~ \rm erg~ s^{-1}$ at $\theta =8\degr$. Our solutions have $\Gamma \simeq 6-10$, at or below this plateau, so we read their Figure 4 at the parameters of each case. Our jet kinetic powers range from 3.8 to 5.5 $\times 10^{45}~ \rm erg~ s^{-1}$, and only Case III falls below the corresponding cold-lepton upper limit of  $\sim 10^{47} ~ \rm erg~ s^{-1}$ from \citet{Mehta2009}.

\citet{Edwards+2006} find the lower limit on the jet bulk motion of $\Gamma>9$ on parsec scales, whereas our kpc-scale solutions give $\Gamma = 9.3$ and 6.1 for Case I and Case III, respectively. We do not regard the lower values as being in conflict with the parsec-scale limit. \citet{Hardcastle2006} found that the bulk Lorentz factors demanded by IC/CMB, $\Gamma \sim 10$, are incompatible with those inferred from kpc-scale jets in lobe-dominated sources, and concluded that jet velocity structure is therefore needed, with the fast material confined to a spine. In the radially stratified models of \citet{Liniewicz2026} the bulk Lorentz factor decreases from its on-axis value to unity at the jet boundary and the emission is integrated over a radius-dependent Doppler factor. A Lorentz factor obtained from a one-zone fit averages across the emitting cross-section and is therefore lower than the on-axis value traced by the parsec-scale motion.

A similar conclusion has already been established for the outer knots (WK 7.8 + WK 8.9 + WK 9.7). \citet{Lucchini+2017} demonstrated that evolving the electron distribution with radiative and adiabatic cooling naturally suppresses the GeV output of the IC/CMB models for the outer knots, allowing fits to the knot and integrated-jet SEDs that remain below Fermi limits for small viewing angles (\(\theta\!\sim\!5^{\circ}\)–\(6^{\circ}\)) and moderate Doppler factors (\(\delta\!\sim\!8\)–11). However, \citet{Meyer2017} emphasized that if the X-rays are dominated by IC/CMB for the outer knots, then once the X-ray normalization is fixed the predicted GeV spectrum is effectively a scaled copy of the synchrotron SED. Using new ALMA measurements to anchor the radio–submillimeter slope, they showed that the \citet{Lucchini+2017} model spectra are steeper than the data and substantially under produce the ALMA fluxes; adopting a steeper than observed synchrotron slope artificially lowers the expected GeV emission for a given X-ray flux and can lead to incorrect conclusions about IC/CMB viability. They further demonstrated that the updated Fermi upper limits rule out a pure IC/CMB origin for the bright outer knots at \(8.7\sigma\) and, assuming equipartition, constrained the large-scale jet to \(\delta<5.3\).

In conclusion, we provide the first comprehensive IC/CMB model for the inner 3\farcs4 to 7\arcsec jet of PKS 0637$-$752. Methodologically, our work is new in two ways. First, we constrain $(\Gamma,\delta,\theta)$ by adopting minimum energy condition in the comoving frame for each trial viewing angle and then sampling $\theta$ from a beaming-aware prior, which yields a self-consistent posterior on orientation and beaming for the inner jet. Second, we combine the ALMA detections with the radio and X-ray fluxes to isolate regimes where a single-population IC/CMB fit would either violate Fermi limits or fail to match the millimeter synchrotron, thereby motivating a mixed (IC/CMB+synchrotron) X-ray origin in the inner jet. Overall, the literature and our analysis indicate that the inner jet favors small viewing angles and mildly to highly relativistic bulk motion on kiloparsec scales, and our work is new in demonstrating that the inner jet X-rays can be partly IC/CMB while remaining consistent with Fermi, with the residual X-rays supplied by a secondary population of synchrotron emitting electrons. The broad-band radiation properties of large scale relativistic jets pose a challenge for particle acceleration processes far from the central black hole.

\newpage

\appendix
\section{Analytical IC/CMB Framework} \label{app:iccmb}
This appendix presents the step-by-step derivation of the IC/CMB model employed in the main text to interpret the kpc-scale X-ray jet of \object{PKS\,0637–752}. We (i) review the synchrotron and inverse-Compton radiation processes, (ii) derive the minimum-energy magnetic field for a cylindrical, synchrotron emitting jet, (iii) apply the necessary relativistic transformations between the jet and observer frames, (iv) obtain an independent magnetic field estimate from the observed radio-to-X-ray flux ratio, which reflects the ratio of magnetic to CMB energy densities, and (v) equate the two magnetic field estimates to solve iteratively for the bulk Lorentz factor, thereby recovering the jet’s rest-frame parameters and kinetic power. Throughout we adopt c.g.s.\ units and the symbol conventions of the main text.

\subsection{Radiation Mechanism of Jets} \label{app:radiation}

Jets radiate predominantly via synchrotron emission at radio–IR wavelengths while inverse-Compton scattering may be important at higher energies.  In the IC/CMB scenario, the X-ray photons arise when the same relativistic electrons that generate the radio synchrotron spectrum up-scatter CMB photons.  The model relies on four assumptions:

\begin{enumerate}
  \item A homogeneous, cylindrical emission region with volume filling factor $\phi=1$, the heavy particle to electron energy ratio k=1, and uniform rest-frame magnetic field~$B$.
  \item A single power-law electron distribution
        $N(\gamma)\,d\gamma=\kappa\,\gamma^{-m}d\gamma$
        between $\gamma_{\min}$ and $\gamma_{\max}$,
        with isotropic pitch-angle distribution.
  \item Thomson-limit scattering of a isotropic CMB with energy density
        $\rho_{\rm cmb}=4\times10^{-13}(1+z)^{4}\ {\rm erg\,cm^{-3}}$;
        synchrotron-self-Compton and other seed fields are negligible on kpc scales.
  \item A steady, one-zone flow: the radio and X-ray photons are co-spatial and share a single bulk Lorentz factor~$\Gamma$ and corresponding Doppler factor~$\delta$.
\end{enumerate}

These assumptions yield analytical relations that translate the observed flux densities, spectral index, and jet volume into the intrinsic power of the jet and other key physical parameters.

\subsection{Synchrotron Emission and the Minimum-Energy Condition} \label{app:minenergy}
The gyro-frequency of an electron in a magnetic field is

\begin{equation}
\nu_{g}=\frac{eB}{2\pi c\,m_{e}},
\label{eq:A1}
\end{equation}
where, $e$ is the charge of electrons in statcoulomb, $B$ is the magnetic field strength in Gauss, $c$ is the speed of light in $\rm cm~ s^{-1}$ and $m_e$ is the mass of the electron in grams. The relativistic electrons with energy $E = \gamma m_e c^2$ emit most of their synchrotron power at the critical frequency, $\nu_c$, defined as
\begin{equation}
\nu_{c}=\frac{3\gamma^{2}eB\sin\theta_{p}}{2\pi c\,m_{e}}.
\label{eq:A2}
\end{equation}
Here, $\theta_{p}$ is the pitch angle i.e., the angle between the velocity of electrons and the magnetic field.
So the synchrotron emission at frequency $\nu$ (Hz) is emitted primarily by electrons with Lorentz factor of
\begin{equation}
\gamma \approx \sqrt{\frac{2 \pi c~ m_{\rm e}}{e B}} \simeq4.886\times10^{-4}\left(\frac{\nu}{B}\right)^{1/2}.
\label{eq:A3}
\end{equation}

For a power-law number distribution of electrons, $N(\gamma) d\gamma = \kappa \gamma^{-m} d\gamma$, the total energy in the electrons ($U_e$) with $\gamma_{min}\leq\gamma \leq \gamma_{max}$ is given as
\begin{equation}
    U_e = \int_{\gamma_{\rm min}}^{\gamma_{\rm max}} E N(\gamma) \,d\gamma \
    = \int_{\gamma_{\rm min}}^{\gamma_{\rm max}} \gamma m_e c^2 \kappa \gamma^{-m} \,d\gamma \
    = \kappa m_e c^2 \int_{\gamma_{\rm min}}^{\gamma_{\rm max}} \gamma^{1-m} \,d\gamma. \
    \label{eq:A4}
\end{equation}

Substituting $m = 2 \alpha + 1$, where $\alpha$ is the power-law slope of the synchrotron emission ($S_{\nu} \propto \nu^{-\alpha}$) and plugging in the constants
\begin{equation}
    U_e  = \kappa m_e c^2 \int_{\gamma_{\rm min}}^{\gamma_{\rm max}} \gamma^{2\alpha} \,d\gamma, \ 
    \label{eq:A5}
\end{equation}
\begin{equation}
    U_e (\rm erg~ cm^{-3}) = \kappa~ m_e c^2  \left ( \frac{\gamma_{\rm max}^{1-2\alpha}}{1-2\alpha} -  \frac{\gamma_{\rm min}^{1-2\alpha}}{1-2\alpha} \right).
    \label{eq:A6}
\end{equation}

This equation holds for $\alpha \neq 0.5$. The monochromatic synchrotron luminosity for a power-law number distribution of electrons satisfying $\gamma_{min}^2 \nu_g\leq\nu \leq \gamma_{max}^2 \nu_g$, is given as
\begin{equation}
     L_{\nu} = 2 \pi \sqrt{3}~ c~ m_e r_e \nu_g \sin \theta_p \int F(\nu, \nu_c) N(\gamma) \,d\gamma \\
    \label{eq:A7}
\end{equation}

\begin{equation}   
     =\kappa \nu^{-(m-1)/2} \nu_g^{(m+1)/2} \sin \theta_p^{(m+1)/2} m_e~ c~ r_e \frac{2 \pi~ 3^{m/2}}{m+1} \Gamma \left(\frac{m}{4}+\frac{19}{12}\right) \Gamma \left(\frac{m}{4}-\frac{1}{12}\right).
    \label{eq:A8}
\end{equation}
Assuming the pitch angle distribution to be isotropic, i.e.,
\begin{equation}
    \sin \theta_p \rightarrow{} \frac{\sqrt{\pi}~ \Gamma \left( \frac{2\alpha +6}{4}\right)}{2 \Gamma\left( \frac{2~ \alpha +8}{4}\right)} = \frac{\sqrt{\pi}~ \Gamma \left[(\alpha+3)/2\right]}{2~ \Gamma \left [ (\alpha+4)/2 \right]}
    \label{eq:A9}
\end{equation}

and substituting $\sin \theta_p$, gyro frequency (equation \ref{eq:A1}) and $m = 2 \alpha + 1$ in equation (\ref{eq:A8}) gives
\begin{eqnarray}
    L_{\nu} & = \kappa \nu^{-\alpha} \left(\frac{e B}{2 \pi c m_e} \right)^{\alpha+1} \frac{\sqrt{\pi}~ \Gamma \left[(\alpha+3)/2\right]}{2~ \Gamma \left [ (\alpha+4)/2 \right]} m_e~ c~ r_e ~ \frac{2 \pi~ 3^{(\alpha+\frac{1}{2})}}{2\alpha+2}~ \Gamma \left(\frac{2\alpha+1}{4}+\frac{19}{12}\right)~ \Gamma \left(\frac{2\alpha+1}{4}-\frac{1}{12}\right)
    \label{eq:A10}
\end{eqnarray}
Collecting the numerical constants and the Gamma function in the factor $C1$, the above equation reduces to
\begin{eqnarray}
 L_{\nu} & = \kappa \nu^{-\alpha} B^{\alpha+1} / C_1,
 \label{eq:A11}
\end{eqnarray}
which can be inverted to 
\begin{equation}
    \kappa = C_1 L_{\nu} \nu^{\alpha} B^{-\alpha-1}.
    \label{eq:A12}
\end{equation}
Substituting $\kappa$ in Equation (\ref{eq:A6}) gives the total energy in electrons.

Next, we assume that the total energy in heavy particles ($U_{\rm{hp}}$) is proportional to the total energy in the electrons, $U_{\rm{hp}} = k U_e$, where, $k$ is the baryon to electron relativistic energy. So the total particle energy density is,
\begin{equation}
    \mathcal{U}_{\rm{part}} =  \frac{U_e + U_{\rm{hp}}}{\phi V} = \frac{U_e + k U_e}{\phi V} = \frac{(1+k) U_e}{\phi V}.
    \label{eq:A13}
\end{equation}
Here, $\phi$ is the volume filling factor and $V$ is the source volume emitting photons observed in the snapshot. Both particle and magnetic energy densities can be expressed as a function of observed quantities and magnetic field strength. Using equation (\ref{eq:A6}) and (\ref{eq:A12}),  total particle energy density simplifies to
\begin{equation}
     \mathcal{U}_{\rm{part}} = \frac{(1+k)}{\phi V} C_1 L_{\nu} \nu^{\alpha} B^{-\alpha-1}~  m_e c^2 \left ( \frac{\gamma_{\rm{max}}^{1-2\alpha}}{1-2\alpha} -  \frac{\gamma_{\rm{min}}^{1-2\alpha}}{1-2\alpha} \right).
     \label{eq:A14}
\end{equation}

The total energy density is the sum of particle and magnetic energy densities
\begin{eqnarray}
    \mathcal{U}_{\rm{tot}} & = \mathcal{U}_{\rm{part}} + \mathcal{U}_{\rm{mag}},
    \label{eq:A15}
\end{eqnarray}
where, the magnetic energy density is given as
\begin{equation}
    \mathcal{U}_{\rm{mag}} = B^2 / 8 \pi.
    \label{eq:A16}
\end{equation}

Next we find the magnetic field strength that minimizes the total energy density.
\begin{eqnarray}
   \frac{ \,d \mathcal{U}_{\rm{tot}}}{\,dB} & = 0, \rm{or}~~ \frac{\,d \mathcal{U}_{\rm{mag}}}{\,dB} = - \frac{\,d \mathcal{U}_{\rm{part}}}{\,dB}\\
    \frac{2 B_{\rm{min}}}{8 \pi} & =  (\alpha +1) B_{\rm{min}}^{-\alpha-2} \left[  \frac{(1+k)}{\phi V} C_1 L_{\nu} \nu^{\alpha}  m_e c^2 \left ( \frac{\gamma_{\rm{max}}^{1-2\alpha}}{1-2\alpha} -  \frac{\gamma_{\rm{min}}^{1-2\alpha}}{1-2\alpha} \right) \right]\\
    B_{\rm{min}}^{\alpha+3} & = 4 \pi (\alpha+1) \left[  \frac{(1+k)}{\phi V} C_1 L_{\nu} \nu^{\alpha}  m_e c^2 \left ( \frac{\gamma_{\rm{max}}^{1-2\alpha}}{1-2\alpha} -  \frac{\gamma_{\rm{min}}^{1-2\alpha}}{1-2\alpha} \right) \right]\\
    B_{\rm{min}} (\rm{Gauss}) & = (4 \pi~ m_e c^2)^{1/(\alpha+3)} \left[  \frac{(1+k)}{\phi V} C_1 L_{\nu} (\alpha + 1) \nu^{\alpha} \left ( \frac{\gamma_{\rm{max}}^{1-2\alpha}}{1-2\alpha} -  \frac{\gamma_{\rm{min}}^{1-2\alpha}}{1-2\alpha} \right) \right]^{1/(\alpha+3)}
    \label{eq:A17}
\end{eqnarray}
So far we have not considered the relativistic transformation of $\nu$, $L_{\nu}$ and $V$.
\subsection{Relativistic transformations} 
$\Gamma$ is the bulk Lorentz factor. Then the bulk velocity of the jet is
\begin{equation}
    \beta = \sqrt{1-\frac{1}{\Gamma^2}}
    \label{eq:A18}
\end{equation}
The Doppler factor is defined as a function of $\beta$ and the angle of line of sight to the jet axis $\theta$
\begin{equation}
    \delta = \frac{1}{\Gamma (1-\beta \cos \theta)} = \frac{1}{\Gamma \left(1-\sqrt{1-\frac{1}{\Gamma^2}} \cos \theta \right)}
    \label{eq:A19}
\end{equation}
The frequency of radiation in the rest frame of the source is $\nu/\delta$. For a cylindrical jet component with measured angular length on the sky as $\theta_l$, an assumed cross jet angle of $\theta_r$ and the unknown angle between the jet and our line of sight be $\theta$, the volume in the source frame is given as
\begin{equation}
    V(\rm{cm^3}) = \frac{\pi \theta_l \theta_r^2}{\delta~ \sin \theta }, 
\end{equation}
and the luminosity is given as
\begin{equation}
    L_{\nu} =  \frac{4 \pi D_{\rm{L}}^2 S_{\nu} (1+z)^{\alpha-1}}{\delta^{3}},
\end{equation}
where, $S_{\nu}$ is the observed flux density at observed frequency $\nu$ and $D_L$ is the luminosity distance

Transforming the minimum magnetic field given by equation (20) in the source frame
\begin{equation}
    B_{\rm{min}}^{\rm source} (\rm{Gauss})  = (4 \pi~ m_e c^2)^{1/(\alpha+3)} \left[  \frac{(1+k)}{\phi V} C_1 L_{\nu} (\alpha + 1) \left (\frac{\nu}{\delta}\right )^{\alpha} \left ( \frac{\gamma_{\rm{max}}^{1-2\alpha}}{1-2\alpha} -  \frac{\gamma_{\rm{min}}^{1-2\alpha}}{1-2\alpha} \right) \right]^{1/(\alpha+3)}
\end{equation}
where, $V$ and $L_{\nu}$ are in the source frame and calculated using equation (23) and (24), respectively.

\subsection{Inverse-Compton of CMB photons by the same electron population}

The power radiated by one synchrotron electron by inverse-Compton scattering of isotropic CMB photon with energy density $\rho_{\rm{cmb}} = a T_{\rm cmb}^4 (1+z)^4 = 4 \times 10^{-13} (1+z)^4 ~\rm{erg~ cm^{-3}}$ is given as 
\begin{equation}
    P_c = \sigma_t c \gamma^2 \rho_{\rm{cmb}} = \frac{8 \pi e^4 \gamma^2 \rho_{\rm{cmb}}}{3 m_e^2 c^3}
\end{equation}
where $\sigma_t$ is the Thompson total cross-section. As for a given particle energy, $E = \gamma m c^2$, the power radiated by inverse-Compton is inversely proportional to the rest mass of the particle by a fourth power ($\propto 1/m^4)$, Therefore, the Compton scattering of CMB photons by heavy particles like proton is negligible. 

The power radiated by the an electron moving freely in a magnetic field $B$ by synchrotron emission is
\begin{equation}
    P_s = \frac{2 e^4 \gamma^2 (B \sin \theta_p)^2}{3 m_e^2 c^3}.
\end{equation}
For electron velocity randomly oriented with respect to the magnetic field, $<\sin^2 \theta_p> = 2/3$, therefore,
\begin{equation}
    P_s = \frac{e^4 \gamma^2 (B \sin \theta_p)^2}{ m_e^2 c^3}.
\end{equation}
From equation (28) and (26),the ratio of the power radiated in synchrotron and Compton process is
\begin{equation}
    \frac{P_s}{P_c} = \frac{B^2 }{8\pi \rho_{\rm{cmb}}}.
\end{equation}
From the observed radio flux density, $S_{\rm{radio}}$ at frequency $\nu_R$ and the observed X-ray flux density $S_{\rm{x}}$ at $\nu_X$, we can determine the magnetic field needed to produce the observed X-ray flux density by Compton scattering of the same synchrotron emitting electron population. Modifying the equation (48) of Felten \& Morrison (1966), the flux density ratio is expressed as
\begin{eqnarray}
    \frac{S_{\rm{radio}}}{S_{\rm{x}}} \left( \frac{\nu_R}{\nu_X} \right)^{\alpha} & = \frac{B^2 }{8\pi \rho_{\rm{cmb}}} \left [ \frac{2 \times 10^4 T_{\rm{cmb}}}{B}\right]^{1-\alpha}
\end{eqnarray}
where, $T_{\rm{cmb}}= 2.725 (1+z)$ is the temperature of the CMB radiation at the redshift of the observed quasar and $\alpha$ is the radio spectral index. We can solve for the magnetic field from the above equation and lets call it $B_{\rm{cmb}}$ and substituting $T_{\rm{cmb}}$, $\rho_{\rm{cmb}}$ 
\begin{eqnarray}
    \frac{S_{\rm{radio}}}{S_{\rm{x}}} \left( \frac{\nu_R}{\nu_X} \right)^{\alpha} & = \frac{B_{\rm{cmb}}^{1+\alpha}}{8\pi \rho_{\rm{cmb}}} \left [ 2 \times 10^4 T_{\rm{cmb}}\right]^{1-\alpha} \\
    B_{\rm{cmb}} & = \frac{[8\pi \times 4 \times 10^{-13} (1+z)^4]^{\frac{1}{1+\alpha}}}{[2 \times 10^4 \times 2.725 (1+z)]^{\frac{1-\alpha}{1+\alpha}}} \left [\frac{S_{\rm{radio}}}{S_{\rm{x}}} \left( \frac{\nu_R}{\nu_X} \right)^{\alpha} \right]^{\frac{1}{1+\alpha}}\\
    B_{\rm{cmb}} & = \frac{[8\pi \times 4 \times 10^{-13}]^{\frac{1}{1+\alpha}}}{[5.45 \times 10^4]^{\frac{1-\alpha}{1+\alpha}}} (1+z)^{\frac{3+\alpha}{1+\alpha}} \left[\frac{S_{\rm{radio}}}{S_{\rm{x}}} \left( \frac{\nu_R}{\nu_X} \right)^{\alpha} \right]^{\frac{1}{1+\alpha}}
\end{eqnarray}
Following \citet{Dermer+Menon2009}, we convert the $B_{\rm{cmb}}$ to the source frame using
\begin{equation}
    B_{\rm{cmb}}^{\rm source} ({\rm{Gauss}}) = \Gamma \left( 1 + \frac{\beta^2}{3} \right)^{1/(1+\alpha)} B_{\rm{cmb}}.
\end{equation}

\begin{acknowledgments}

We thank the anonymous referee for their constructive feedback. A.S. and D.A.S. acknowledge support from NASA Contract NAS8-03060 to the {\sl Chandra X-ray Center}. J.M. has been supported by Chandra grants  GO1-22086X and GO2-23097X from the {\sl Chandra X-ray Center}. J.M. is grateful to Axel Donath and Diab Jerius for helpful guidance on installing SAOTrace.

This research has made use of data obtained from the Chandra Data Archive, and software 
provided by the Chandra X-ray Center (CXC) in the application packages CIAO \citep{Fruscione2006,Fruscione2026} and Sherpa\citep{Freeman2001,Siemiginowska2024}.

This paper employs a list of Chandra datasets, obtained by the Chandra X-ray Observatory, contained in the Chandra Data Collection (CDC) 277 ~{\href{https://doi.org/10.25574/cdc.277}{doi:10.25574/cdc.277}}
and 411 ~{\href{https://doi.org/10.25574/cdc.411}{doi:10.25574/cdc.411}}.
The National Radio Astronomy Observatory is a facility of the National Science Foundation operated under a cooperative agreement by Associated Universities, Inc. 
This paper makes use of the following ALMA data: ADS/JAO.ALMA\#2012.1.00075.S, 2012.1.00335.S, 2012.1.00554.S, 2012.1.00603.S, 2012.1.00641.S, 2012.1.01108.S, 2013.1.00063.S, 2013.1.00070.S, 2013.1.00280.S, 2013.1.00287.S, 2013.1.00450.S, 2013.1.00535.S, 2013.1.00700.S, 2013.1.00993.S, 2015.1.00190.S, 2015.1.00204.S, 2015.1.00979.S. ALMA is a partnership of ESO (representing its member states), NSF (USA) and NINS (Japan), together with NRC (Canada), NSTC and ASIAA (Taiwan), and KASI (Republic of Korea), in cooperation with the Republic of Chile. The Joint ALMA Observatory is operated by ESO, AUI/NRAO and NAOJ. The authors thank the ALMACAL team for sharing the data used in this work. 

This research is based on observations made with the NASA/ESA Hubble Space Telescope obtained from the Space Telescope Science Institute, which is operated by the Association of Universities for Research in Astronomy, Inc., under NASA contract NAS 5–26555. These observations are associated with program GO 14696
~{\href{https://doi.org/10.17909/js73-am83}{doi:10.17909/js73-am83}}.
\end{acknowledgments}

%






\bibliography{references}
\bibliographystyle{aasjournalv7}



\end{document}